\documentclass[twocolumn]{aastex63}
\usepackage{ifpdf}

\defcitealias{Cantat-Gaudin_2018}{CG18}
\defcitealias{Ferraro_2012}{F12}
\defcitealias{Ahumada_2007}{AL07}
\defcitealias{Rain_2020}{R20}
\defcitealias{Vaidya_2020}{V20}

\usepackage{color}

\shorttitle{Blue straggler stars in Trumpler~5, Trumpler~20, and NGC~2477}
\shortauthors{Rain et al.}
\graphicspath{{./}{figures/}}

\begin{document}

\title{\Large The blue straggler population of the open clusters\\ Trumpler~5, Trumpler~20, and NGC~2477}

\author[0000-0003-4009-8316]{Rain, M.~J.}

\author[0000-0002-0155-9434]{Carraro, G.}
\affiliation{Dipartimento di Fisica e Astronomia, Universit\`{a} di Padova, Vicolo Osservatorio 3, I-35122, Padova, Italy}

\author[0000-0002-7091-5025]{Ahumada, J.~A.}
\affiliation{Observatorio Astron\'{o}mico, Universidad Nacional de C\'{o}rdoba, Laprida 854, X5000BGR, C\'{o}rdoba, Argentina}

\author[0000-0001-6205-1493]{Villanova, S.}
\affiliation{Departamento de Astronom\'{\i}a, Universidad de Concepci\'{o}n, 160 Casilla, Concepci\'{o}n, Chile}

\author[0000-0002-9486-4840]{Boffin, H.}
\affiliation{ESO, Karl-Schwarzschild Strasse 2, D-85748 Garching, Germany}

\author[0000-0002-6555-0842]{Monaco, L.}
\affiliation{Departamento de Ciencias Fisicas, Universidad Andres Bello, Santiago de Chile, Chile}

\begin{abstract}

We present a study, based on \emph{Gaia}~DR2, of the population of blue straggler stars (BSS)  in the open clusters Trumpler~5, Trumpler~20, and NGC~2477.  All candidates were selected according to their position in the color-magnitude diagram, to their proper motion components, and to their parallax. We also looked for yellow stragglers, i.e., possible evolved blue stragglers.
We found that Trumpler~5 hosts a large BSS population, which allowed us to analyze their radial distribution as a probe of the cluster's dynamical status. The BSS distribution was compared with that of red giant branch stars (RGB) to evaluate mass segregation. Our results indicate that BSS are not more centrally concentrated than RGB stars in any of the clusters. The radial distribution of BSS in Trumpler~5 is flat. Additionally, using a multi-epoch radial velocity survey conducted with the high-resolution spectrograph FLAMES/GIRAFFE at  VLT, we measured the radial velocities of a sample of stragglers, for the sake of comparison with the mean radial velocity and velocity dispersion of the clusters. Based on the radial velocity variations for different epochs, we roughly classified these stars as possible close- or long-period binaries.

\end{abstract}

\keywords{star clusters and associations: general --- 
star clusters and associations: individual: Trumpler~5, Trumpler~20, NGC~2477 --- blue stragglers --- binaries: close}

\section{Introduction}\label{sec:introduccion} 

\begin{table*}
\centering
	\caption{Main parameters of the  open clusters under study}
	\label{tab:clusters_parameters}
	\begin{tabular}{l c c c c c c c c c c c c } 
		\hline
  Cluster &  $l$ & $b$ & Distance & $E(B-V)$ &  $R_{a}^{1}$ &$\log(\mathrm{age})$ & [Fe/H] & V$_{R}$ \\
         &[deg] & [deg] & [kpc] & [mag]  & [arcmin] & [yr] & [dex] & [km~s$^{-1}$]  \\
\hline
 Trumpler~5  &   202.86 & $+$1.05      & 3.19$^{b}$ & 0.62$^{a}$ & 15.4$^{a}$ & 9.60$^{a}$ & $-0.40\pm0.006^{a}$ & $+49.67\pm0.66^{c}$  \\
 Trumpler~20 & 301.47 & $+$2.22     & 3.56$^{b}$ & 0.46$^{a}$ &  16.0$^{a}$ & 9.11$^{a}$ & $+0.17\pm0.030^{a}$ & $-40.94\pm1.20^{e}$  \\
 NGC~2477    & 253.56 & $-5.64$  & 1.44$^{b}$ & 0.31$^{a}$ &  15.0$^{a}$ & 8.85$^{a}$ & $+0.07\pm0.030^{a}$ & $~+08.62\pm0.46^{d}$  \\

		\hline
	\end{tabular}\\
	$^{1}$Apparent radius\\
	$^{a}$\citet{Dias_2014}
	$^{b}$\citet{Cantat-Gaudin_2018}
	$^{c}$\citet{Monaco_2014}
	$^{d}$\citet{Mishenina_2015}
	$^{e}$\citet{Carraro_2014-2}
\end{table*}

Blue Straggler stars (BSS) are among the large variety of exotic objects that populate stellar clusters. These stars are located above and bluewards the turnoff (TO) in the color-magnitude diagram (CMD) of globular clusters (e.g \citealt{Sandage_1953, Piotto_2004}), dwarf galaxies (e.g \citealt{Momany_2015}), open clusters (e.g \citealt{Milone_1994, Ahumada_1995, Ahumada_2007}), and even of field populations of the Milky Way \cite{Preston_2000, Santucci_2015}. Two different  mechanisms have been proposed to make a BSS: mass exchange in a binary system, and the collision of two stars induced by stellar interactions in a dense stellar environment. The first scenario suggests that BSS are formed from  primordial binaries that evolve mainly in isolation, until  mass transfer starts, eventually leading up to a  possible merger \citep{McCrea_1964}. The second scenario indicates that BSS are the product of a dynamical merger between single stars or binary systems \citep{Hills_1976, Davies_1994}. Nevertheless,  these objects are extremely important because they can provide information on the dynamics, the binary population, and the history of stellar evolution of the cluster they belong to
\citep{Bailyn_1995, Ferraro_2009, Chen_2009, Glebbeek_2010, Wyse_2020}. An extensive review of their properties has been compiled and presented in \cite{Boffin_2015}

Despite their presence in all stellar environments, 
there is an increasing interest in studying the blue straggler population in open clusters. A few catalogs of BSS  in OCs are currently available, but their main disadvantage  is the lack of reliable membership information, as they are based on purely photometric selection criteria  \citep[hereafter AL95 and AL07]{Ahumada_1995, Ahumada_2007}. While useful, these compilations are not reliable enough to allow the derivation of statistical properties of BSS because, unfortunately, field stars tend to occupy the very same region as the stragglers  in the CMD \citep{Carraro_2010}. An improvement in the selection of BSS  has become possible thanks to the second \emph{Gaia} Data Release (DR2) \citep{Gaiacoll2018},  that permits a better discrimination of genuine BSS from field stars using high-quality proper motion and parallax information.

BSS are strongly affected by dynamical friction, mainly because of their masses---around 1.2--1.7~$M_{\odot}$, according to 
observations; therefore, they are perfect test particles to probe the impact of dynamics in different stellar systems. 
The  observed radial distribution of BSS in globular clusters (GCs) has been systematically studied.  In this context, \citet[hereafter F12]{Ferraro_2012} grouped GCs in three families based on their BSS radial distributions. Clusters of Family~I, or dynamically young GCs, show a flat distribution; in these systems the dynamical friction has not yet caused visible effects, even in the innermost regions. In Family~II GCs, the dynamical friction has become more efficient and the mass segregation has started, leading to a bimodal distribution with a peak in the innermost region, followed by a  minimum at $r_{\mathrm{min}}$; there is a rise (or maximum) of the BSS density in the outer regions, where the stars  have  still  not  been  affected  by  the  dynamical friction. Finally, in the Family~III class of GCs, the external maximum disappears, and the only noticeable peak in the distribution is the central one, followed by a minimum. A similar work in open clusters has been carried out by \cite{Vaidya_2020}. The authors determined accurate stellar membership and studied the BS radial distribution on seven open clusters. Five of them (Melotte~66, NGC~2158, NGC~2506, NGC~188, and NGC~6791) were assigned to Family~II. In the remaining two clusters (Berkeley~39 and NGC~6819) they found flat distributions. Recently, individual clusters have also been explored in this context. \cite{Bhattacharya_2019}  studied the radial distribution of the very old  cluster Berkeley~17 ($\sim10$~Gyr, \citealt{Kaluzny_1994}), and placed it into the Family~II class of GCs;  \citet[hereafter R20]{Rain_2020} showed that Collinder~261 ($\sim9$~Gyr, \citealt{Dias_2014}) has a flat BSS distribution, like that of Family~I clusters.

\begin{figure*}
   \includegraphics[width=\linewidth]{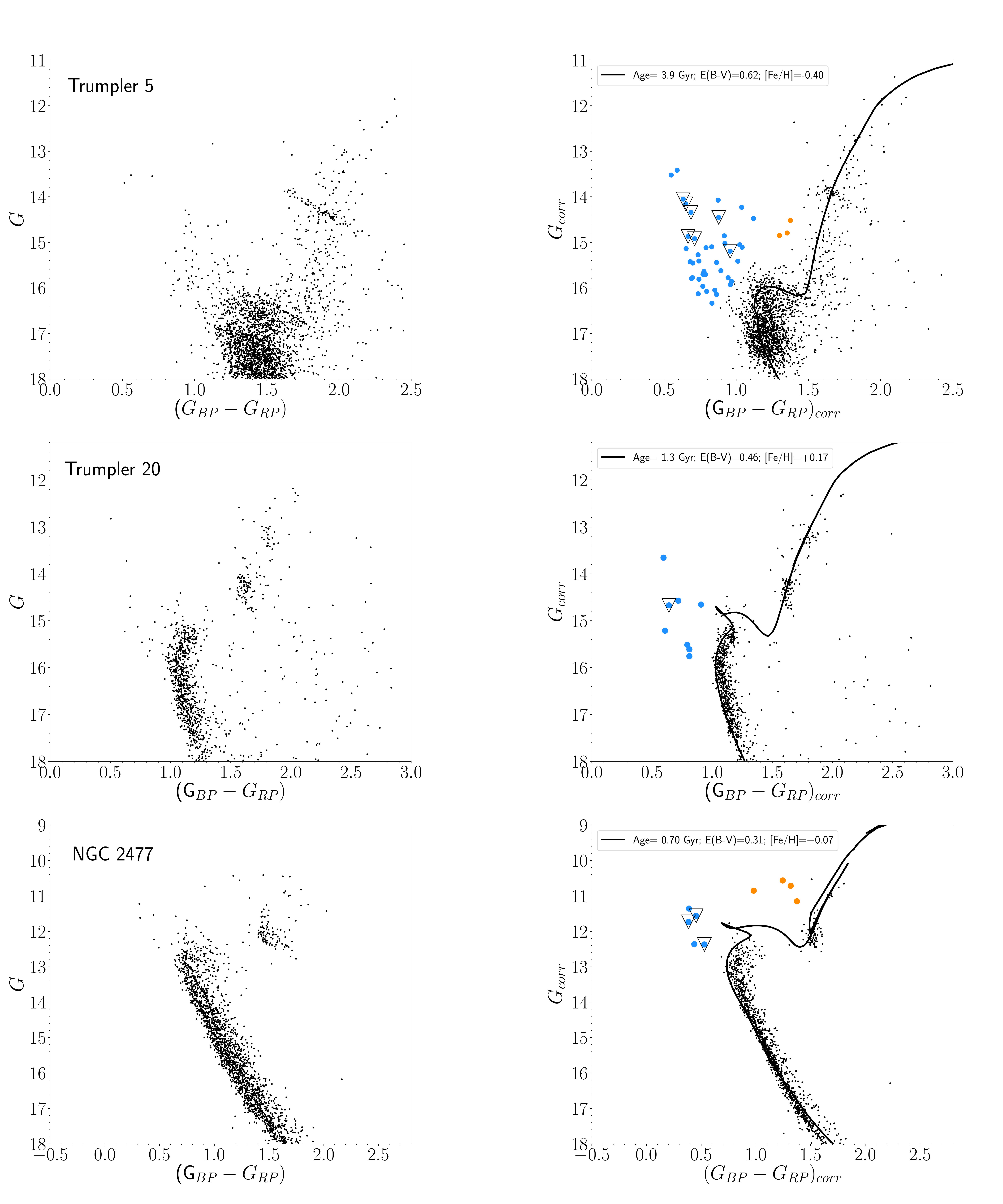}
    \caption{\textbf{Left panels:} Color-magnitude diagrams of Trumpler~5, Trumpler~20, and NGC 2477. Black dots are cluster members with $P_{\mathrm{memb}} \geq 50$\%, selected by \citet{Cantat-Gaudin_2018}. \textbf{Right panels:} Color-magnitude diagrams after the correction for  DR, for stars with $P_{\mathrm{memb}} \geq 50$\%, and falling within their corresponding radius $R$ (see \S~\ref{subsec:field_contamination}). Blue solid dots are our BSS candidates. Orange filled circles are yellow straggler stars. Black open triangles are the sources with spectroscopic data available in this paper, cross-matched with BSS candidates of \citetalias{Ahumada_2007} only for Trumpler~5 and NGC~2477, since Trumpler~20 was not included in the 2007 catalog.
    Isochrones are from \cite{Bressan_2012}.}
    \label{fig:cmd}
\end{figure*}

While the photometric studies of globular and open clusters are often cited as an introduction to blue stragglers, there are not many spectroscopic studies that explore this population, specially in open clusters.  NGC~188 and M67 are the main laboratories for detailed spectroscopic studies of their BSS populations, since both have very well identified BSS members (e.g., \citealt{Peterson_1984, Mathieu_2009, Mathieu_2015, Geller_2011, Geller_2012, Milliman_2013}). Besides these two clusters, spectroscopic studies  in open clusters are available only for individual stars, for example:  NGC~6791 \citep{Brograard_2018};  NGC~6087,  NGC~6530, and Collinder~223 \citep{Aidelman_2018}; NGC~2141 \citep{Luo_2015}; and Collinder~261 (\citetalias{Rain_2020}). In a statistical analysis of the distribution of the masses of the secondaries, \cite{Geller_2011}  found that the companions of the BSS of NGC~188 have masses values of $\sim$ 0.5 $M_{\odot}$, suggesting possible white dwarf (WD) companions. In fact, such WD companions have been detected for seven BSS in NGC~188 using far-ultraviolet HST observations \citep{Gosnell_2015}. Currently, and thanks to the works of \cite{Geller_2009} and \cite{Geller_2012}, we know that the percentage of binaries among BSS is significantly larger ($\sim 75\%$) than in the main sequence (MS) of OCs, where it is of about 20\%,  and also that the orbital period distribution of BSS is quite different from that of MS binaries, with the majority of BSS having orbital periods close to 1000~days, and most of them likely including a WD companion. 

In this paper we study the BSS population of three open clusters: Trumpler~5, Trumpler~20, and NGC~2477. These clusters cover a wide range of ages, metallicities, and positions in the Milky Way (see Table~\ref{tab:clusters_parameters}).

Trumpler~5 (C0634$+$094, $\alpha=06^\mathrm{h}36^\mathrm{m}29^\mathrm{s}$, $\delta=+09^\circ28'12''$, J2000.0),  discovered and cataloged by \cite{Trumpler_1930}, is an old ($\sim$ 4 Gyr), metal-poor, massive, and very populous open cluster. Trumpler~20 (C1236$-$603, $\alpha=12^\mathrm{h}39^\mathrm{m}45^\mathrm{s}$, $\delta=-60^\circ38'06''$, J2000.0)  resides in the inner disk and is located  beyond the great Carina spiral  arm; it is  old ($\sim$ 1.3 Gyr), metal-poor, and not so distant  \citep{Platais_2008}. Finally, NGC~2477 (C0750$-$384, $\alpha=07^\mathrm{h}52^\mathrm{m}10^\mathrm{s}$, $\delta=-38^\circ31'48''$, J2000.0) is an intermediate-age open cluster ($\sim$ 0.7 Gyr) with near-solar metallicity.

The layout of the paper is as follows. In Section~\ref{sec:datasets} we present the  datasets  used.   In  Section~\ref{sec:phot_analysis} we describe the photometric analysis and the selection criteria of blue straggler  stars in open clusters. We also explore the radial density profile of each cluster and estimate the field contamination. Finally we give the results of our selection in the three clusters. In Section~\ref{sec:spec-analysis} we explain how the spectra were reduced,  the radial velocities were extracted, and their uncertainties   estimated; in this section we also define the criteria to establish membership and binary status of our targets, and  discuss the results of the spectroscopic  detections. Section~\ref{sec:radial_density_profiles} is entirely devoted to explore the BSS radial distribution of Trumpler~5  and, finally, in Section~\ref{sec:conclusions} we give a  summary of this study and the conclusions.

\section{Datasets} \label{sec:datasets}
\subsection{Photometric data} \label{subsec:photometric_data}

We used the Data Release~2  of the 
European Space Agency mission \emph{Gaia}\footnote{\url{https://gea.esac.esa.int/archive/}} \citep{gaia16,gaia18}. For more than a billion stars, 
this survey provides a five-parameter astrometric solution: position, trigonometric parallax, and proper 
motion, as well as photometry in three broad-band filters ($G$, $G_\mathrm{BP}$, and $G_\mathrm{RP}$). The \emph{Gaia} catalog also gives radial velocities for about 7 million stars,  mostly brigther
than $G \sim 13$. The astrometric solution,  the photometric contents and validation, and
the properties and validation of radial velocities are described in
\citet{Lindegren_2018}, \citet{Evans_2018}, and \citet{Katz_2019}, respectively.

\subsubsection{Differential Reddening}
\label{sec:dr}

\begin{figure}
\setlength{\lineskip}{0pt}
\centering

   \includegraphics[width=0.8\linewidth]{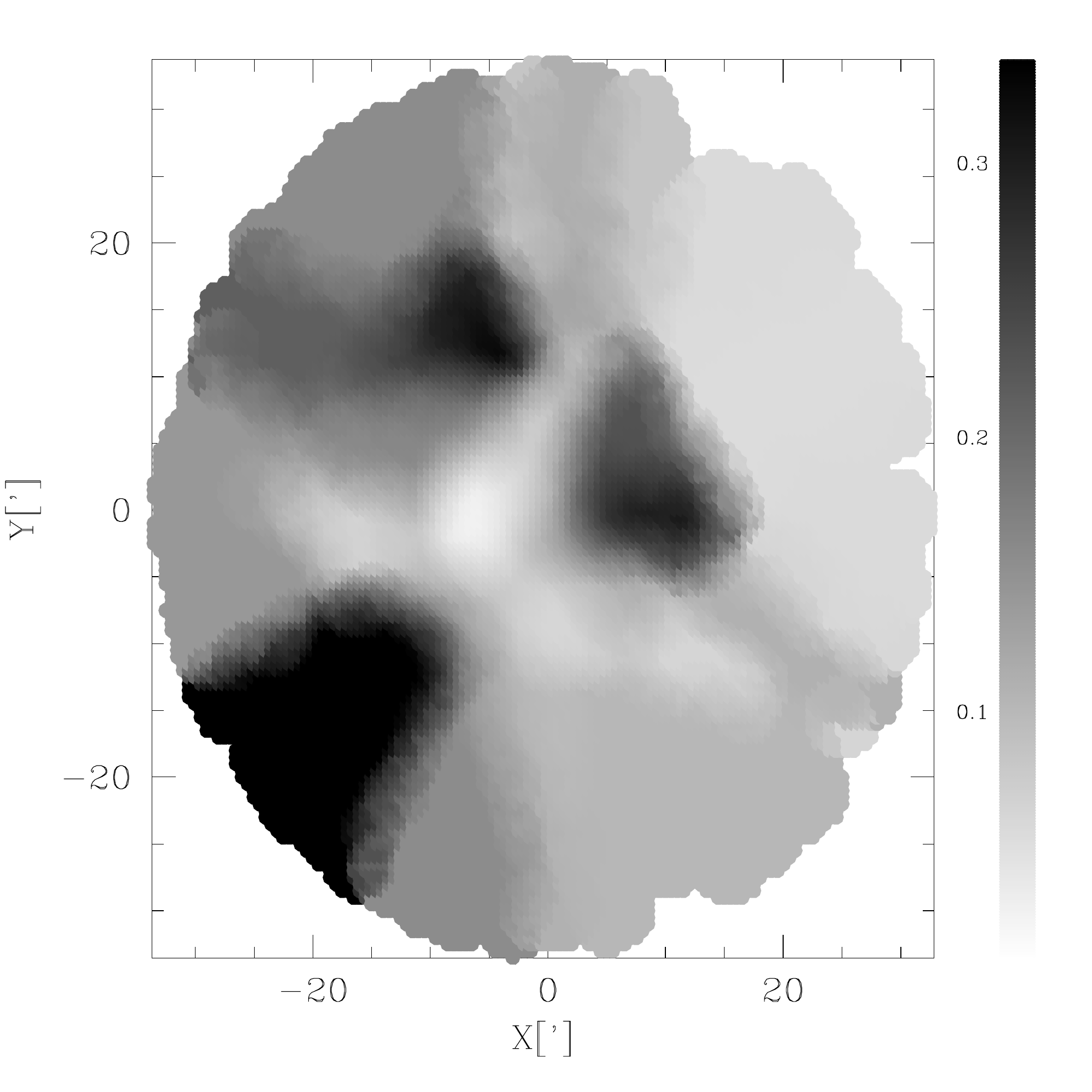}
   
   \includegraphics[width=0.8\linewidth]{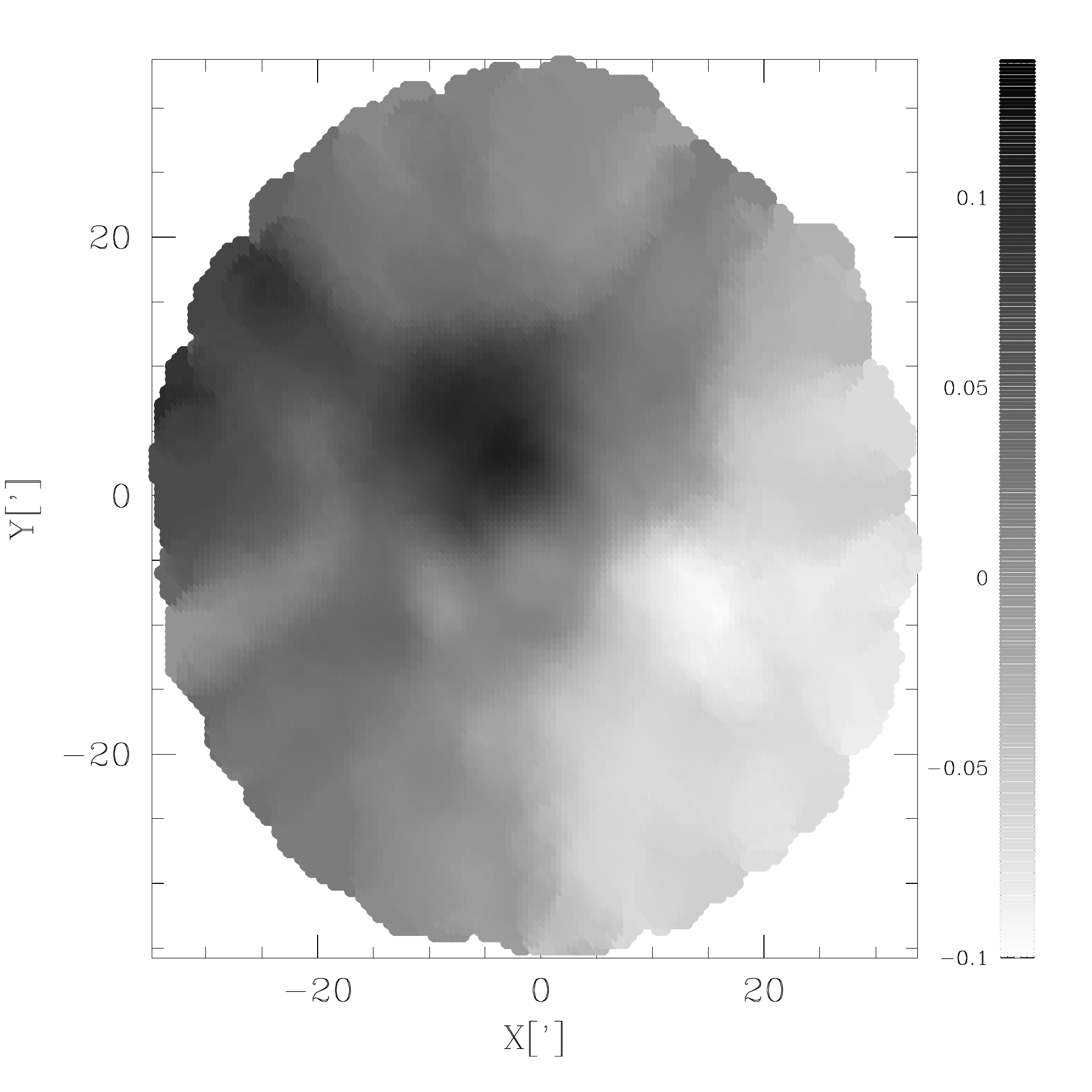}

   \includegraphics[width=0.8\linewidth]{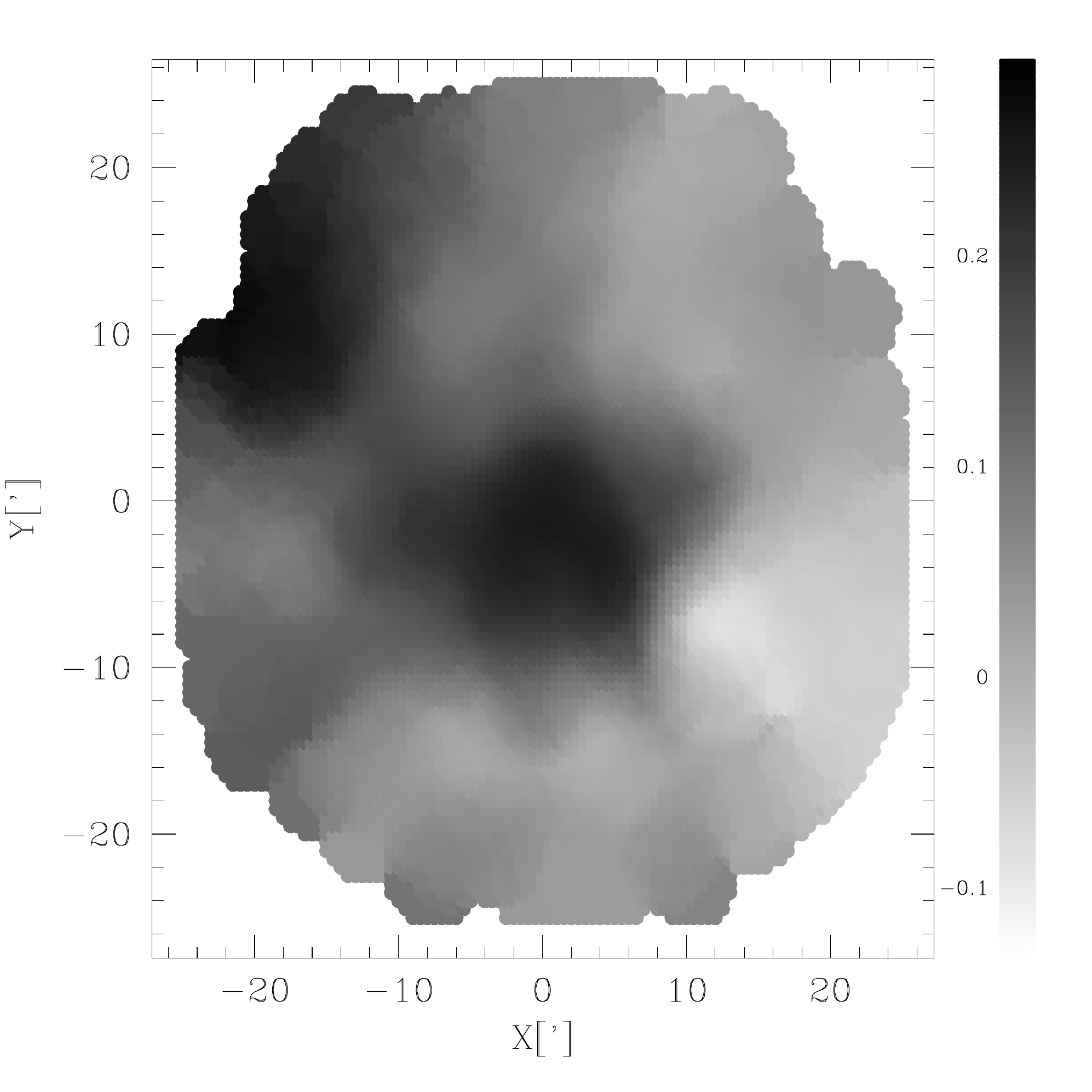}

\caption{Reddening maps in RA ($X$) and Dec ($Y$) (expressed in arcmin) obtained for Trumpler~5, Trumpler~20, and NGC~2477 
(from top to bottom). The intensity code indicates where differential reddening is stronger (darker) or weaker (lighter).}
    \label{fig:reddening_maps}
\end{figure}

\begin{table}
\centering
	\caption{Details of the spectroscopic observations for Trumpler 20 (February 11--12  to March 3--4, 2012,  and  January 30--31 to March 28--29, 2018); Trumpler 5 (February 11--12 and 29  to March 28--29, 2018); and NGC 2477 (October 28--29, 2011 to March 8--9, 2012, and January 30--31 to March 3--4, 2018). All dates are  Modified Julian Date (JD$-$2,400,000.5)}
	\label{tab:obs_details}
	\setlength\tabcolsep{2pt}
	\begin{tabular}{c c c c} 
		Trumpler 20 &  & &\\
		\hline
				Epoch  &  MJD$_{\mathrm{start}}$  & MJD$_{\mathrm{end}}$ & Exposure \\
		 & [d] & [d] & [s] \\
		\hline
	    1 & 55969.306622551 & 55969.3344003785  & 2400.0043\\
		2 & 55990.200693457 & 55990.2284712834	& 2400.0042 \\
		3 & 58149.292577118 & 58149.3203549282  & 2400.0028\\
		4 & 58206.118721651 & 58206.1464994681  & 2400.0034\\
		\hline
    \end{tabular}
    \setlength\tabcolsep{2pt}
	\begin{tabular}{c c c c} 
		Trumpler 5 & & &\\
		\hline
		Epoch  &  MJD$_{\mathrm{start}}$  & MJD$_{\mathrm{end}}$ & Exposure \\
		 & [d] & [d] & [s] \\
		\hline
		1 & 55969.166866965 & 55969.1923299766	& 2400.0042\\
	    2 & 55987.039632022 & 55987.0650950232  & 2200.0033\\
		3 & 58177.077345320  & 58177.1028083142 & 2200.0027\\
		4 & 58180.077972559 & 58180.1034357268  & 2200.0177\\
		5 & 58206.000367976 & 58206.0258309702  & 2200.0027\\
		\hline
	\end{tabular}
	\setlength\tabcolsep{2pt}
	\begin{tabular}{cccc} 
	  	NGC 2477 & & &\\
		\hline
		Epoch  &  MJD$_{\mathrm{start}}$  & MJD$_{\mathrm{end}}$ & Exposure \\
		 & [d] & [d] & [s] \\
		\hline
	    1 & 55863.327186374 & 55863.3445475059  & 1500.0018\\
		2 & 55863.345536336 & 55863.3628974807	& 1500.0029\\
		3 & 55995.200983276 & 55995.2183444288  & 1500.0036\\
		4 & 58179.147028199 & 58179.1643893541  & 1500.0038\\
		5 & 58181.169961441 & 58181.1873225822  & 1500.0026\\
        6 & 58122.199733802 & 58122.2170949270  & 1500.0012\\
    	7 & 58135.279591133 & 58135.2969522545  & 1500.0009\\
		8 & 58149.265756268 & 558149.2831174243 & 1500.0039\\
		\hline
	\end{tabular}
\end{table}

The main effect of differential reddening (DR) in a CMD is a broadness (or dispersion) of the sequences of the cluster; this results from  the differential presence of dust along the line of sight and across the field of view, causing different extinction values \citep{Platais_2008}. For old open clusters ($\log(\mathrm{age})\geq 1$~Gyr) the effects of DR are most noticeable in the TO and the RGB morphologies. In this context, the position of Trumpler~5 and Trumpler~20 in the Galactic disk and their high values of reddening (see Table~\ref{tab:clusters_parameters}) suggest these two clusters are highly affected by DR. This is noticeable in the left panels of Figure~\ref{fig:cmd}, where the elongated red clump and the thick appearance of the main sequence and TO in Trumpler~5 are clear indicators of DR. In minor scale, the same effects are observed in the CMD of Trumpler~20. We will quantify, however, the effect of DR in the three clusters analysed in this work.
The DR correction was performed in two different ways. In the case of Trumpler~5, which is the cluster most affected 
by this problem, we miss most of its MS due to the cut in $G = 18$ (see \citetalias{Cantat-Gaudin_2018} for more details). 
On the other hand, its CMD shows a well defined RC dispersed along the reddening line. First of all we selected RC stars 
and determined the reddening law $R_{G}=A_{G}/E(G_\mathrm{BP}-G_\mathrm{RP})$ by a linear least square fit. We got $R_{G}=1.79\pm 0.05$. 
Then we selected an arbitrary point along the RC line as the zero-point for the reddening correction. This point has $G=13.90$ and $(G_{\mathrm{BP}}
-G_\mathrm{RP})=1.67$. Then, for each RC star we calculated the distance, both vertical and horizontal, with respect to the reference point. The vertical distance gives the differential $A_{G}$ absorption at the position of the star, while the horizontal distance gives the differential $E(G_\mathrm{BP}-G_\mathrm{RP})$ 
reddening at the position of the star. After this first step, for each star of the field (both cluster and non-cluster members) we selected the three nearest RC stars and calculated the mean differential $A_{G}$ absorption and the mean differential reddening $E(G_\mathrm{BP}-G_\mathrm{RP})$, and finally subtracted this mean value from the star's ($G_\mathrm{BP}-G_\mathrm{RP}$) color and  $G$ magnitude.

For Trumpler~20 and NGC~2477, we used instead  MS stars for the DR correction, since we have for them much longer and populated main sequences (see Fig.~\ref{fig:cmd}).
 We defined a line along the MS, and for each one of the selected MS stars, we calculated its distance from this line along the reddening law line. For this reddening law, the line was assumed the slope we found for Trumpler~5, i.e., $R_{G}=1.79$. The vertical projection of this distance gives the differential $A_{G}$ absorption at the position of the star, while the horizontal projection gives the differential reddening $E(G_\mathrm{BP}-G_\mathrm{RP})$ at the star's position. After this first step, for each star of the field (both cluster members and non-members) we selected the ten nearest MS stars and calculated the mean differential $A_{G}$ absorption and the mean differential $E(G_\mathrm{BP}-G_\mathrm{RP})$, and finally subtracted this mean value to its  ($G_\mathrm{BP}-G_\mathrm{RP}$) color and $G$ magnitude.

We want to underline the fact that the number of reference stars used for the reddening correction (three for Trumpler~5 and ten for Trumpler~20 and NGC~2477), is a compromise between having a correction affected as little as possible by photometric random errors, and the highest possible spatial resolution. 
Figure~\ref{fig:cmd} shows the uncorrected
(left) and corrected (right) color-magnitude diagrams. Figure \ref{fig:reddening_maps}, finally, shows the reddening maps for each cluster.

\begin{figure}
\setlength{\lineskip}{0pt}
\centering
\vspace{-0.30cm}   
       
   \includegraphics[width=0.83\linewidth]{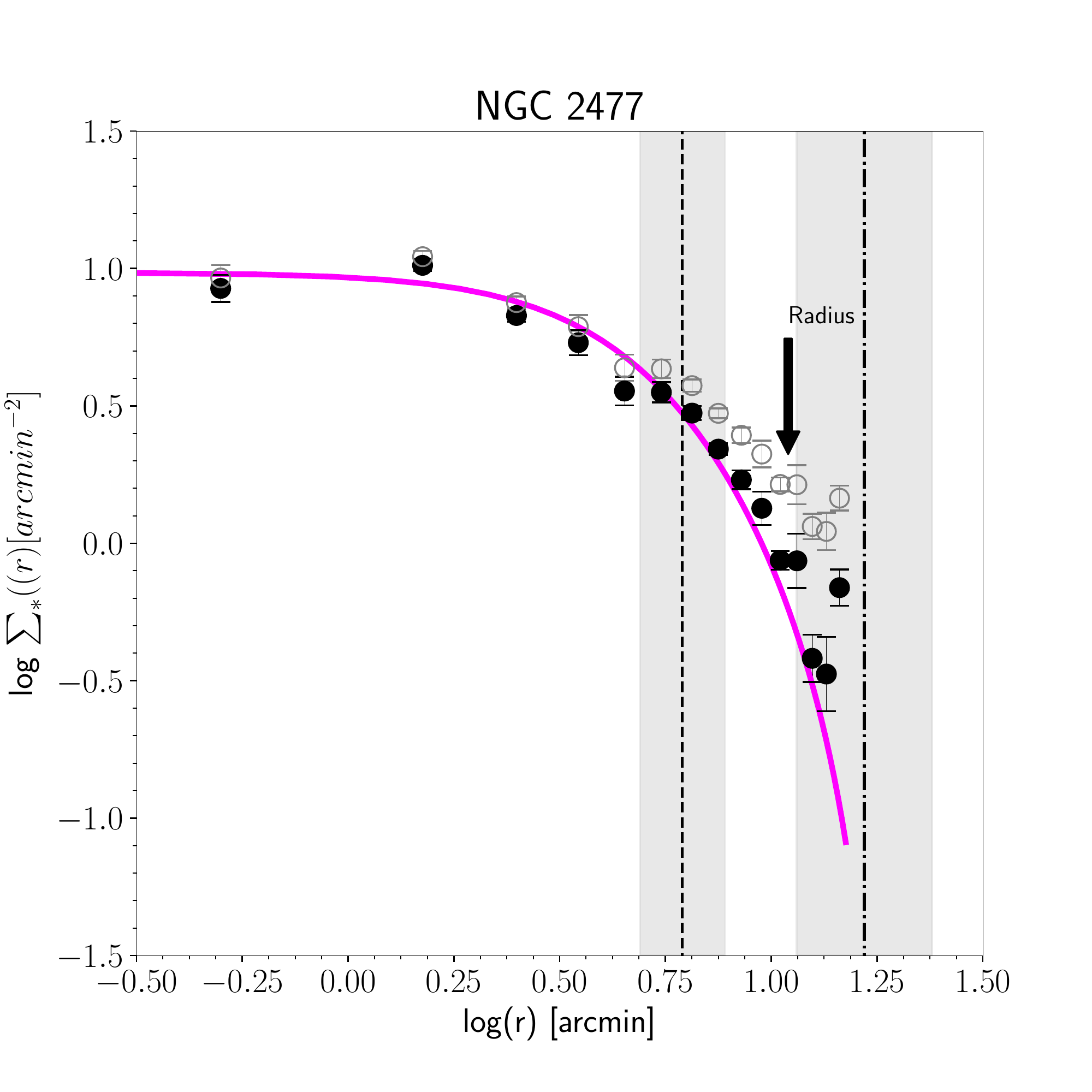}
   
\vspace{-0.30cm}   
   
   \includegraphics[width=0.83\linewidth]{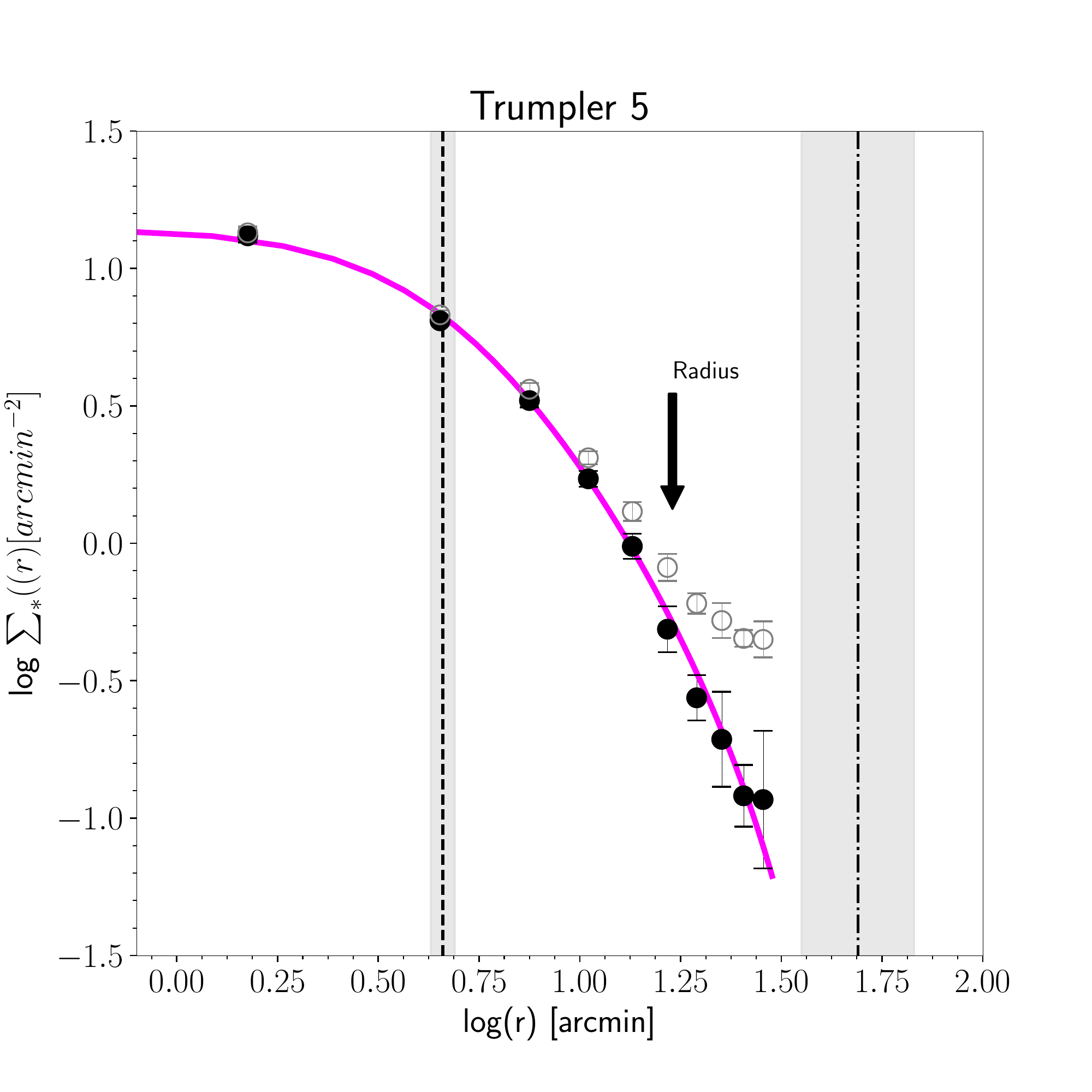}
   
\vspace{-0.30cm}   
      \includegraphics[width=0.83\linewidth]{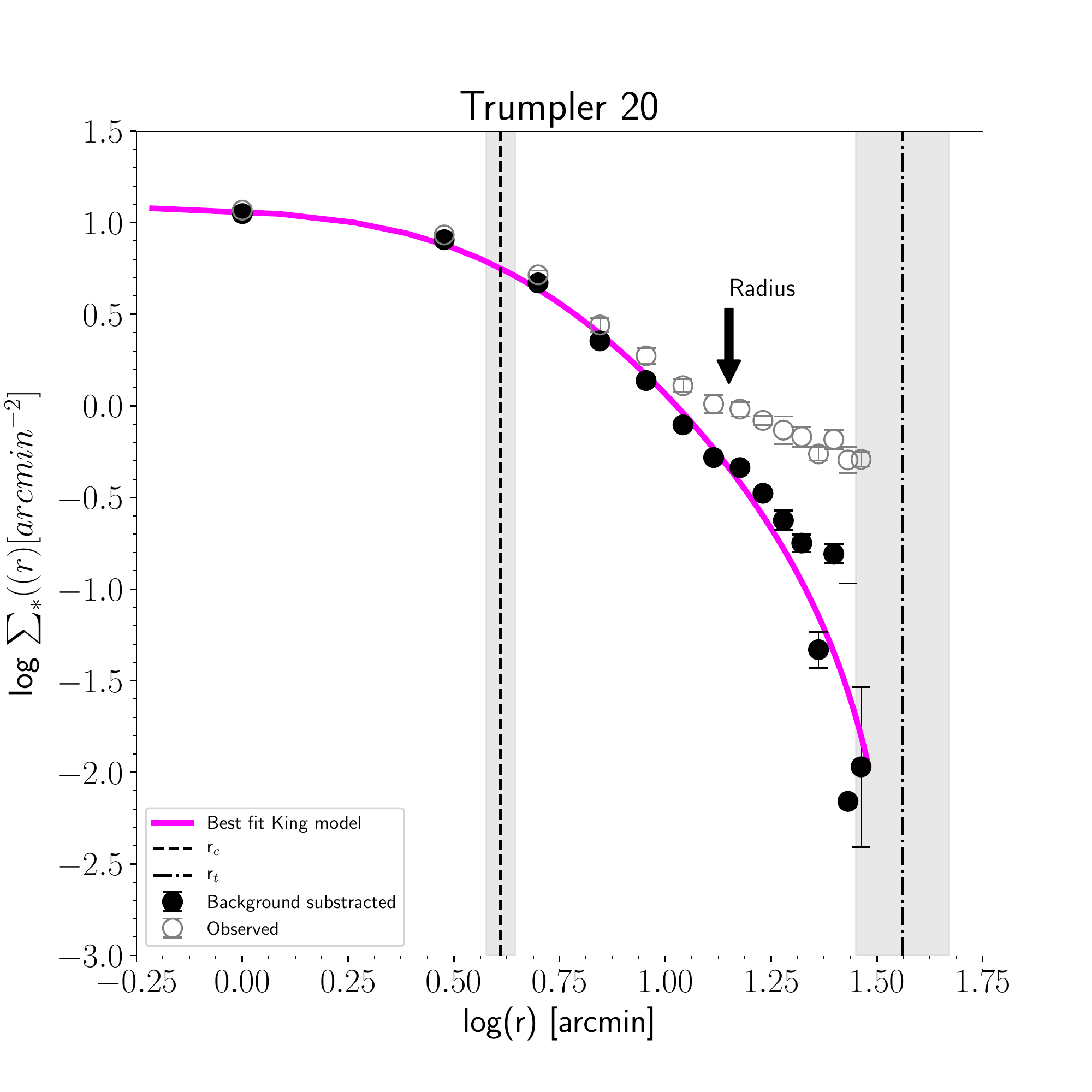}
   
\vspace{-0.5cm}   
  
\caption{Observed density profiles of NGC~2477, Trumpler~5, and Trumpler~20 (gray open circles). The background subtracted profile (black filled circles) is  the difference between the observed profile and the background level. The magenta line is the best fit King model. The dashed and dash-dotted vertical lines correspond to the values of the core and tidal radius, respectively, with their  uncertainties plotted as grey shaded regions. The corresponding radius R is indicated for every cluster on each panel.}
    \label{fig:kingmodels}
\end{figure}

\subsection{Spectroscopic data} \label{subsec:spectroscopic_data}
The clusters were observed with the Fibre Large Array Multi Element Spectrograph (FLAMES)\footnote{\url{http://www.eso.org/sci/facilities/paranal/instruments/flames.html}}
attached to the Very Large Telescope (VLT) of the European Southern Observatory  (ESO; Paranal Observatory, Chile), using the combination of the mid-resolution spectrograph GIRAFFE and the fiber link to UVES.   NGC~2477 data were collected in two periods,
 October~2011 to March~2012,   and January~2018 to March~2018. Data for clusters Trumpler~5 and Trumpler~20   were also obtained in two periods,   February~2012 to March~2012,  and January~2018 to March~2018. These data were gathered under ESO programs 088.D-0045(A) and 0100.D-0052(A). 

The UVES fibers were allocated to the cluster's clump stars, whose membership is very solid, to set the zero point of the radial velocity. 
The reduction and  analysis of the UVES data are described in \citet{Carraro_2014-1, Carraro_2014-2, Monaco_2014, Mishenina_2015}. GIRAFFE was used with the setup HR8, which covers the wavelength range 491.7--516.3~nm, with a spectral resolution  $R \equiv \lambda/\Delta\lambda\equiv$~20,000. Integration time ranged between 1,500~sec and 2,400~sec depending on the cluster. In total, NGC~2477 was observed on eight epochs, Trumpler~5 on five epochs, and Trumpler~20 on four epochs. Some details of the observations are given in Table~\ref{tab:obs_details}. For the GIRAFFE data we only performed the sky-subtraction and normalization  using the IRAF\footnote{IRAF is distributed by the National Optical Astronomy Observatory, which is operated by the Association of Universities for Research in Astronomy, Inc., under cooperative agreement with the National Science Foundation.} packages \texttt{sarith} and \texttt{continuum}, since they had  already been reduced in Phase~3\footnote{\url{http://archive.eso.org/wdb/wdb/adp/phase3_spectral/form}}.

\section{Photometric Analysis}\label{sec:phot_analysis}

We took advantage of the selection of cluster members already performed by \citet[hereafter CG18]{Cantat-Gaudin_2018}, who used the membership assignment code UPMASK\footnote{Unsupervised  Photometric Membership Assignment in Stellar  Clusters} \citep{Krone_2014}. UPMASK depends on minimal physical assumptions about the stellar cluster region, and assigns the membership probabilities on the basis that member stars are more concentrated than a random distribution (field stars), and that share properties such as proper motions and parallaxes. \citetalias{Cantat-Gaudin_2018} selected data from \emph{Gaia}~DR2 considering as members those stars  located over a radius twice the value  reported by \citet[hereafter DAML02]{Dias_2002}, with proper motions within 2~mas~yr$^{-1}$, and with parallaxes within 0.3 mas, of those of the cluster centroid ($\mu_{\alpha}\cos\delta$, $\mu_{\delta}$, $\varpi$).

\subsection{Cluster mean proper motions and parallaxes}

For consistency, we carried out the same photometric analysis described in our previous paper (\citetalias{Rain_2020}). To use the membership selection of \citetalias{Cantat-Gaudin_2018} with confidence, we calculated the mean proper motions and parallaxes values for the three clusters.

As mentioned in \S~\ref{subsec:photometric_data}, in \emph{Gaia}~DR2 the radial velocity information is provided for targets with  $G<13$~mag. Here, we identify these objects in all three clusters. First, we selected giant stars with V$_{R}$ measurements available and we calculated a first estimation of the mean V$_{R}$ value and the corresponding rms for each cluster. Then, only stars within V$_{R} \pm \sigma$ were selected. The same procedure was performed on each cluster until reach considerable small values of the rms---of the order of the errors in \emph{Gaia} radial velocities---and V$_{R}$ values similar to those reported in Table \ref{tab:clusters_parameters}. We ended up with 10, 38, and 79 stars for Trumpler~5, Trumpler~20, and NGC~2477, respectively.


Second, we searched \emph{Gaia} counterparts of members previously identified in the literature with confirmed membership. For Trumpler~5 we used five red-clump stars identified as members by \cite{Monaco_2014}. In the case of Trumpler~20, we used a sample of five bona~fide clump stars for which \cite{Carraro_2014-2} determined the abundance of several elements and their ratios to confirm their membership. Finally, for NGC~2477 we used six giant stars analyued and identified as members by \citet{Bragaglia_2008}. We cross-correlated the position on the sky of these stars and the \emph{Gaia} DR2 catalog, looking for the nearest neighbours within 1$''$.  Finally, using both stars from the literature and the ones selected above we calculated the mean proper motions and parallaxes values. The values we found are reported in Table \ref{tab:proper_motions}. Errors indicated are the standard deviations from the stars.  
 
Our results are in remarkable agreement with the values in the literature \citep[e.g.,][]{Cantat-Gaudin_2018, Gao_2018}. However, all of them differ considerably from the values reported in DAML02. The most extreme case is that of  Trumpler~20, with absolute differences (our work \emph{minus} DAML02) of  4.87 mas yr$^{-1}$ and 3.89 mas yr$^{-1}$ for  $\mu_{\delta}$ and $\mu_{\alpha}\cos\delta$, respectively.  These differences are unlikely to be caused by systematic errors in \emph{Gaia} data, and probably arise from significant contamination by field stars and the lack of reliable cluster membership in the DAML02 catalog. For Trumpler~5 and NGC~2477 we found differences of 0.24 mas yr$^{-1}$ and 0.42 mas yr$^{-1}$, and 1.84  mas yr$^{-1}$ and 0.94 mas yr$^{-1}$ for $\mu_{\delta}$ and $\mu_{\alpha}\cos\delta$, respectively. The values we estimated were not used in the rest of this analysis, but we always employed those of  \citetalias{Cantat-Gaudin_2018}.

\subsection{King profiles and structural parameters}

\label{subsec:king_profile}

For each  cluster we determined the stellar density profiles and derived the structural parameters such  as the core and tidal radius and concentration parameters (Table \ref{tab:structural_parameters}), using a King profile fitting approach \citep{King_1962}. A possible link between cluster dynamical quantities and the BSS population is explored in   \S~\ref{sec:radial_density_profiles}.

We constructed the radial density profile of the cluster following \cite{Salinas_2012}. We first divided all observed cluster members into concentric annuli;  then, each annulus was in turn divided into eight sub-sectors defined by wedges of 45$^{\circ}$  centred on the cluster. The cluster center values here used are the ones reported by \citetalias{Cantat-Gaudin_2018}, obtained using stars with $P_{\mathrm{memb}} \geq 50$\%. The density in each sub-sector was measured as the ratio between the number of stars within the sub-sector and the area of the sub-sector itself; this allowed us to obtain a mean stellar surface density and its uncertainty at the mid-point of each shell. The resulting profile was fitted with an isotropic single-mass  \cite{King_1962} model:
\begin{equation}
    n(r)= k \left(\frac{1}{\sqrt{1+(r/r_{c})^{2}}} - \frac{1}{\sqrt{1+(r_{t}/r_{c})^{2}}} \right)^{\!2}\! + b,
\end{equation}
were $r_c$ is the core radius, $r_{t}$ is the tidal radius, $k$ is a scaling constant, and $b$ is the background level. This last one was fixed by the non-linear least-squares method fitted to the total (cluster + background) density profile---i.e., we approximate the model by a linear one and the parameters were estimated by successive iterations---, the fitting gave us $k$,  $r_{c}$, $r_{t}$, and $b$ for each cluster.
After obtaining the cluster parameters, another cluster center was found using the new value of $r_{t}$. With this new center, a new calculation of the radial profile was performed, resulting in a new  $r_{t}$.  The procedure was repeated until the position of the cluster center and the value of $r_{t}$ stopped changing. 

The final density profile for each cluster is shown in  Figure \ref{fig:kingmodels}. The open circles represent the observed density profile, while the black circles mark the background-subtracted profile, obtained as the difference between the observed profile and the background level. The structural parameters are given in Table~\ref{tab:structural_parameters}, where $c=\log(r_{t}/r_{c})$ is the concentration parameter. The derived parameters for Trumpler~20  and NGC~2477 are in good agreement with those measured by \cite{Donati_2014} and \cite{Eigenbrod_2004}, respectively.

\begin{deluxetable*}{l c c c c c c c c c c c c c c c c } 
\tablecaption{Cluster structural parameters from the King profile fittings and derived quantities.\label{tab:structural_parameters}}
\tablehead{
\colhead{Cluster} &  \colhead{$N_{\ast}$} & \colhead{$r_{c}^a$} & \colhead{$r_{t}^b$} &  \colhead{$R$} & \colhead{$c^c$} & \colhead{$t_{\mathrm{relax}}$} & \colhead{$N_{\mathrm{relax}}$} & age \\
 & [arcmin] & [arcmin] & [arcmin] & [arcmin] & & [Myr] & & [Gyr]}
\startdata
 Trumpler~5  & 1908  & $4.57 \pm 1.07$ & $48.97 \pm 15.80$ & 15 & $1.02\pm 0.02$ & 124 & 31.45 & 3.9\\
 Trumpler~20 & 850  & $4.07 \pm 1.01$ & $36.30 \pm 9.27$ & 13 & $0.95\pm 0.02$ & 115  & 10.41 & 1.2\\
 NGC~2477    & 1367  & $6.21  \pm 1.59$ & $16.90  \pm 6.33$ & 12 & $0.43 \pm 0.10$ & 130 & 5.38 & 0.7\\
\enddata
\tablenotetext{}{$^a$~{Core radius} \\
$^b$~{Tidal radius}\\
$^c$~{Concentration parameter}}	
\end{deluxetable*}

\begin{deluxetable*}{l c c c c c c c c c c c c  } 
\tablecaption{Proper motions and parallaxes values.\label{tab:proper_motions}}
\tablehead{
\colhead{Cluster} &  \colhead{$\mu_{\alpha} \cos \delta^a$}  &  \colhead{$\mu_{\delta}^a$} &   \colhead{$\varpi^a$} & \colhead{$\mu_{\alpha} \cos \delta^b$}  &  \colhead{$\mu_{\delta}^b$} &   \colhead{$\varpi^b$}  \\
 & [mas yr$^{-1}$] & [mas yr$^{-1}$] & [mas] & [mas yr$^{-1}$] & [mas yr$^{-1}$] & [mas]}
\startdata
 Trumpler~5  &  $-$0.59  $\pm$ 0.19 & 0.27 $\pm$ 0.18 & 0.28 $\pm$ 0.09 & $-$0.53  $\pm$ 0.18 & 0.24 $\pm$ 0.09  & 0.30 $\pm$ 0.07 \\
 Trumpler~20 &  $-$7.08  $\pm$ 0.09 & 0.18 $\pm$ 0.08 & 0.25 $\pm$ 0.04 & $-$7.09  $\pm$ 0.09 & 0.17 $\pm$ 0.04 &  0.24 $\pm$ 0.02 \\
 NGC~2477    &  $-$2.44  $\pm$ 0.16 & 0.87 $\pm$ 0.19 & 0.66 $\pm$ 0.03 & $-$2.42  $\pm$ 0.17 & 0.88 $\pm$ 0.18 &  0.65 $\pm$ 0.03   \\
\enddata
\tablenotetext{}{$^a$~{\cite{Cantat-Gaudin_2018}\\
$^b$~{This work}}}	
\end{deluxetable*}

\subsection{Identification of the stragglers} \label{subssec:identification_BSS}
\label{sec:identificacion}

The region on the colour-magnitude diagrams where blue stragglers can be found is very well defined. To the left, by the zero-age main sequence (ZAMS); to the right, by the turnoff color; and down, by the magnitude at which the observed sequence of the cluster separates from the ZAMS. We superimposed an isochrone for each cluster and a ZAMS of solar metallicity from \citet{Bressan_2012} to the CMD. The fitting parameters are indicated in each panel of  Figure \ref{fig:cmd}. Additionally, we constrained this region by plotting the equal-mass binary locus (dashed line)  obtained by shifting the isochrones by 0.753~mag in  $G$  toward brighter magnitudes; in this way, we expect that binaries containing normal main-sequence TO stars are excluded. These stars may appear as stragglers, but their components may not be such---see, e.g., \citet{Hurley_1998} for a discussion. A
 red limit in $(B-V)$ was defined in \citetalias{Ahumada_2007} for Trumpler~5 only. Therefore, and as we did for Collinder~261 (\citetalias{Rain_2020}),  to impose the same limit in the \emph{Gaia} system we used the relation of \citet{Jordi_2010}:
\begin{equation}
C_{1} = 0.0187 + 1.6814\, C_{2} - 0.3357\, C_{2}^{2} + 0.117\, C_{2}^{3}.
\end{equation} 
Adopting $C_{2}\equiv (B-V)=0.87$---the limit set in \citetalias{Ahumada_2007}---we found  $C_{1} \equiv G_\mathrm{BP}-G_\mathrm{RP}= 1.30$.  
We used the color of the TO to define the red limit for NGC~2477 and Trumpler~20, since
\citetalias{Ahumada_2007} did not give it for the first cluster, and did not include the second cluster in their catalog.

Stars redder than this limit will be considered as possible yellow straggler stars  (YSS).

Finally, recalling  that the mass-transfer theory for the BSS \citep{McCrea_1964}
sets an upper limit of 2.5 magnitudes above the TO, we will regard stars
above it as potential massive blue stragglers, as \citetalias{Ahumada_2007} did.

\subsection{Field contamination}
\label{subsec:field_contamination}

In the colour-magnitude diagrams of open clusters, the different evolutionary status of the stars and their corresponding sequences---i.e., main sequence, binary sequence, turnoff point, sub-giant and giant branch and, in the case of the older clusters, the red clump---are usually well defined, and provide essential information on the physical properties of the stars which define them.  These sequences are deformed by differential reddening (\S~\ref{sec:dr}) and field contamination; these effects are particularly  noticeable in clusters located towards the Galactic center. In this context, the decontamination of field stars is very important in the study of open clusters. Given the proximity to the Galactic plane of Trumpler~5 and Trumpler~20, and since we identified our BSS and YSS within a radius twice the apparent radius reported in DAML02, we expected field contamination from young stars that---as described in \cite{Carraro_2008}---would occupy similar positions in the CMD as our straggler candidates. Our method for assessing membership using proper motions and parallaxes from \emph{Gaia}~DR2  decreases, but does  not remove all the contamination by such field young stars. To do so, and following \cite{Vaidya_2020}, we limited each cluster's radius to that at which there are more cluster stars than field stars. We estimated the cluster size as the radius \textit{R} where the cluster density profile separates appreciably from that of the background 
(see Figure~\ref{fig:kingmodels}); the  radii obtained this way are  in Table~\ref{tab:structural_parameters}. The rate of false positives was estimated by counting stars located inside an annular region---selected outside the tidal radius---with the same area as that of the cluster. All the selected sources are within $\mu_{\alpha}\cos\delta \pm \sigma$, $\mu_{\delta} \pm \sigma$, and $\varpi \pm \sigma$. 

We found 13 and 2 interlopers for Trumpler~5 and Trumpler~20, respectively. No contamination by field stars was observed in the case of NGC~2477. 


\subsection{Final detections} \label{sec:photometric_detections}

Until now, the most extensive, published  catalog of BSS in open clusters is  \citetalias{Ahumada_2007}. They looked for BSS candidates in the colour-magnitude diagrams of galactic OCs taking advantage of the Open Cluster Database \emph{WEBDA}\footnote{\url{https://webda.physics.muni.cz/}} \citep{Mermilliod_Pauzen_2003} and the Lund Catalogue of Open Cluster Parameters  \citep[ed.\ 1987]{lyn83}, with limited membership information.  The \citetalias{Ahumada_2007}  catalog lists 1887 BSS candidates in 427  clusters. The present work  uses instead the  powerful astrometric solution of \emph{Gaia}~DR2 to identify cluster members, as described  in \S~\ref{sec:phot_analysis}. First, the stragglers were identified over the entire  extension of each cluster, as given by \citetalias{Cantat-Gaudin_2018}. Then, each BS population was limited to the  radius $R$  estimated in \S~\ref{subsec:field_contamination}. Finally, for the three clusters we left out all stars with a membership probability ($P_{\mathrm{memb}}$) below 50\%. This operative and conservative cut-off has been defined in the literature as the probability of containing the most likely members \citep{Cantat-Gaudin_2018, Carrera_2019, Banks_2019, Yontan_2019}. Using the above-mentioned restrictions we thus defined {\it bona~fide},  non-spurious BSS populations. Our results for each cluster are presented in Table~\ref{tab:BS_resume}.

\begin{deluxetable}{l c c c c c c c c c c c c c c c c } 
\tablecaption{Blue and yellow straggler populations after each selection process. \label{tab:BS_resume}}
\tablehead{
\colhead{Cluster} &  \colhead{$N_{\mathrm{BS}}^{a}$} &  \colhead{$N_{\mathrm{BS}}^{b}$ } &  \colhead{$N_\mathrm{{BS}}^{c}$} &  \colhead{$N_\mathrm{{YS}}^{c}$} &  \colhead{$N_{\mathrm{YS}}^{b}$ } &  \colhead{$N_\mathrm{{YS}}^{c}$}}
\startdata
 Trumpler~5  &   177 & 51 & 40 & 7 & 3  & 3  \\
 Trumpler~20 &    65 &  15 & 8 & 0  & 0  & 0  \\
 NGC~2477    &    5  &  5  & 5 &  4  &   4  &  4  \\
\enddata
\tablenotetext{}{$^a$~{Blue/Yellow stragglers over the entire extension of \cite{Cantat-Gaudin_2018}}\\ $^b$~{Blue/Yellow stragglers within the cluster radius $R$ calculated in \S~\ref{subsec:field_contamination}}\\ $^c$~{Blue/Yellow stragglers within the cluster radius $R$  and with $P_{\mathrm{memb}}\geq 50$\%}}	
\end{deluxetable}

Trumpler~5 is the cluster that hosts the largest population of BSS of our sample (see upper panels of Figure~\ref{fig:cmd});  two stars are located above the upper limit of 2.5 magnitudes defined for massive stragglers (\S~\ref{sec:identificacion}). In the case of Trumpler~20, all stars visible in the CMD are within $\sim 1$~mas~yr$^{-1}$, and only one massive straggler candidate was identified. Finally,  NGC~2477 does not harbor many BSS, and no massive stragglers are visible in its CMD.

Additionally, we compared our BSS candidates with those of \citetalias{Ahumada_2007} in Trumpler~5 and NGC~2477 only, since Trumpler~20 was not included in \citetalias{Ahumada_2007}. To do so, we searched the \emph{Gaia} counterparts of the \citetalias{Ahumada_2007} candidates. 
For Trumpler~5, there are only seven BSS in common, and the rest are non-members.   This is not surprising, given the position of this cluster at a low galactic latitude, and therefore expected to suffer from significant field contamination, which  was not accounted for in \citetalias{Ahumada_2007}. In the case of NGC~2477, all BSS identified in \citetalias{Ahumada_2007} are indeed members, but they are all concentrated around the cluster TO; among them,  only three are in our list of BSS candidates. 

\begin{table*}
\centering
 \caption{Individual radial velocity measurements for blue stragglers in Trumpler~5. Binary classification according to their radial velocity variability and rotational velocities is reported in the last two rows.}
 \setlength\tabcolsep{2pt}
 \label{tab:rv_blue_Tr5}
 \begin{tabular}{c c c c c c c c c } 
 \hline 
   &   Tr5-BS1  &  Tr5-BS2  & Tr5-BS3  & Tr5-BS4 & Tr5-BS5 &  Tr5-BS6  & Tr5-BS7 \\
 \hline
  RV$_{1}$ & $+22.72\pm11.12$  & $+46.03\pm3.65$  & $+31.25\pm10.41$   & $+31.71\pm13.75$ & $+03.87\pm7.93$   & $+26.92\pm12.25$ &  $+43.33\pm12.77$   & \\
  RV$_{2}$ & $+24.87\pm6.68$   & $+165.46\pm6.18$ & $+67.26\pm13.90$   & $+26.34\pm14.31$ & $-09.56\pm14.74$  & $+15.73\pm11.25$ &   $+49.34\pm8.01$   &\\
  RV$_{3}$ & $+21.41\pm11.10$  & $-13.43\pm3.16$  & $+25.84\pm12.41$   & $+33.09\pm15.33$ & $-16.44\pm6.35$   & $+13.40\pm12.77$ &   $+09.99\pm6.29$   &\\
  RV$_{4}$ & $+12.03\pm11.96$  & $-22.52\pm3.69$   & $+57.84\pm18.19$    & $+33.23\pm13.56$ & $-13.80\pm12.75$  & $+00.58\pm14.49$ &   $+23.49\pm5.44$ &  \\
  RV$_{5}$ & $+17.80\pm7.70$   & $-17.48\pm3.10$   & $-7.62\pm 16.13$    & $+24.28\pm15.29$ & $-14.77\pm12.95$  & $+10.75\pm14.70$ &   $+04.03\pm4.14$ & \\
  \hline
  Class $^a$ & NM? &  M, CB  & M, CB & NM? & NM, LP? & M, CB? & M, CB & \\
  \hline
  $ v\sin i$ (km~s$^{-1}$)& 150 & 30 & 270 & 170 & 100 & 100 & 160 & \\
  \hline
 \end{tabular}\\
  $^a$ \textbf{NM}: non-member, \textbf{M}: member, \textbf{CB}: close-binary system, \textbf{LP}: long-period binary.\\
\end{table*}

\begin{table}
\caption{Same as in Table \ref{tab:rv_blue_Tr5},  for NGC~2477.}
\setlength\tabcolsep{2pt}
\centering
 \label{tab:rv_blue_NGC2477}
\begin{tabular}{c c c c } 
 \hline 
    & NGC2477-BS1 &  NGC2477-BS2 & NGC2477-BS3 \\
 \hline
 RV$_{1}$ & $+07.78\pm1.63$ & $+11.83\pm2.64$ & $+06.51\pm1.98$ \\
 RV$_{2}$ & $+07.79\pm1.85$ & $+11.80\pm2.80$ & $+06.27\pm2.21$ \\
 RV$_{3}$ & $+08.67\pm1.61$ & $+13.01\pm2.30$ & $+07.49\pm1.85$ \\ 
 RV$_{4}$ & $+08.26\pm1.70$ & $+11.68\pm2.00$ & $+06.18\pm2.10$ \\ 
 RV$_{5}$ & $+08.30\pm1.57$ & $+11.32\pm2.14$ & $+06.76\pm1.87$ \\
 RV$_{6}$ & $+08.53\pm1.55$ & $+10.92\pm1.98$ & $+05.84\pm1.89$ \\
 RV$_{7}$ & $+08.50\pm1.79$ & $+11.63\pm1.87$ & $+06.12\pm1.97$ \\
 RV$_{8}$ & $+08.06\pm1.65$ & $+10.16\pm2.27$ & $+06.96\pm1.88$ \\
   \hline
  Class$^a$ & M & M & M \\
  \hline
  $v\sin i$ (km~s$^{-1}$) & 30 & 45 & 40\\
  \hline 
  \end{tabular}\\
  $^a$ \textbf{M}: member
\end{table}

\begin{table}
\caption{Same as in Table \ref{tab:rv_blue_Tr5},  for Trumpler~20.}
\setlength\tabcolsep{2pt}
\label{tab:rv_blue_Tr20}
\begin{tabular}{l c c c } 
 \hline 
& Tr20-BS1  & Tr20-BS2 & Tr20-BS3 \\
 \hline
  RV$_{1}$ & $+14.81 \pm 2.91$  & $-23.63 \pm 1.16$ & $-07.22 \pm 4.73$ \\
  RV$_{2}$ & $-56.68 \pm 3.98$  & $-26.36 \pm 2.04$ & $-07.97 \pm 4.06$ \\
  RV$_{3}$ & $+22.40 \pm 2.57$  & $-00.27 \pm 2.78$ & $+07.67 \pm 2.88$ \\ 
  RV$_{4}$ & $-57.29 \pm 3.90$  & $+41.52 \pm 2.21$ & $-08.65 \pm 1.95$ \\
  \hline
  Class$^a$ & M, CB &  M, CB &  NM, LP?\\
  \hline
  $v\sin i$  (km~s$^{-1}$)& 40 & 50 & 90\\
  \hline
\end{tabular} \\
\centering
  $^a$\textbf{NM}: non-member, \textbf{M}: member, \textbf{CB}: close-binary system, \textbf{LP}: long-period binary
\end{table}

\section{Spectroscopic Analysis} \label{sec:spec-analysis}

This is the first high-resolution spectroscopic analysis of the BSS population in the open clusters Trumpler~5, Trumpler~20, and NGC~2477.  Unfortunately, not all the candidates were observed with FLAMES, because when the observational time was allocated, we used the BSS list of \citetalias{Ahumada_2007} to select the targets, a list very different from that found in this work using \emph{Gaia} information.

The spectroscopic analysis was carried out on seven out of the 40 blue stragglers in our list for Trumpler~5---plus one star with $P_{\mathrm{memb}} \leq 50$\%, on one out of our eight blue stragglers for Trumpler~20---and two stars not identified as BS in this study, given their low $P_{\mathrm{memb}}$, and in three out of the five blue stragglers in NGC~2477.

\subsection{Radial and rotational velocities } 
\label{subsec:rv}

The radial velocities were calculated with the cross-correlation task IRAF \texttt{fxcor} \citep{Tonry_Davis_1979}. Each spectrum was cross-correlated with synthetic templates  obtained with the SPECTRUM  code\footnote{\url{http://www.appstate.edu/~grayro/spectrum/spectrum.html}} \citep{Gray_Corbally_1994}. The synthetic spectrum was computed with a molecular data file \texttt{stdatom.dat} \citet{Grevesse_1998}, a linelist \texttt{luke.lst}, suitable for mid-B- to K-type stars, and a model atmosphere calculated with the code ATLAS9 \citep{Castelli_Kurucz_2003}.

Among the most interesting observational features of BSS is the projected rotational velocity of $v~sin(i)$. From the theoretical point of view, BSS driven from the proposed formation mechanisms (i.e mass transfer and collisions) are expected to rotate fast. In practice, however, from observations in stellar cluster BSS have been identified as low and fast-rotators (e.g \citealt{Lovisi_2010, Lovisi_2013}). In this sense braking mechanism has been suggested to occur and slow down the stars \cite{Sills_2005}. Given this complex scenario, the selection of a proper template for each star was mandatory given the different rotational velocities the targets have. The model atmospheres were calculated with parameters for F-type MS, or slightly evolved stars ($T_{\mathrm{eff}}=7500$~K and $\log g=4.0$), adopting solar metallicity. The micro-turbulence was set as $\xi = 0.0$~km~s$^{-1}$ for all  templates. Spectra were convolved with a Gaussian  to model the instrumental resolution of the spectrograph, and rotational broadening was applied. Spectra were modelled assuming a $v\sin i$ value varying from 10 to 300 km~s$^{-1}$, with a step of 10 km~s$^{-1}$. For each target, synthetic spectra were then re-normalised to maintain the continuum level at 1 and to match the core-depth of the observed spectral lines. In this way, it was particularly easy to estimate the rotational velocity of the stars just comparing by eye the width of the observed and synthetic spectral lines.  The error of this procedure depends critically on the S/N of the spectra. However, it was never above $\pm 10$~km~s$^{-1}$. 

The radial velocities measurements for the blue population are reported in Table \ref{tab:rv_blue_Tr5}, \ref{tab:rv_blue_Tr20}, and \ref{tab:rv_blue_NGC2477} for Trumpler~5, Trumpler~20, and NGC~2477, respectively.

\subsection{Errors}\label{subsec:errors}
\begin{figure}
\includegraphics[width=\linewidth]{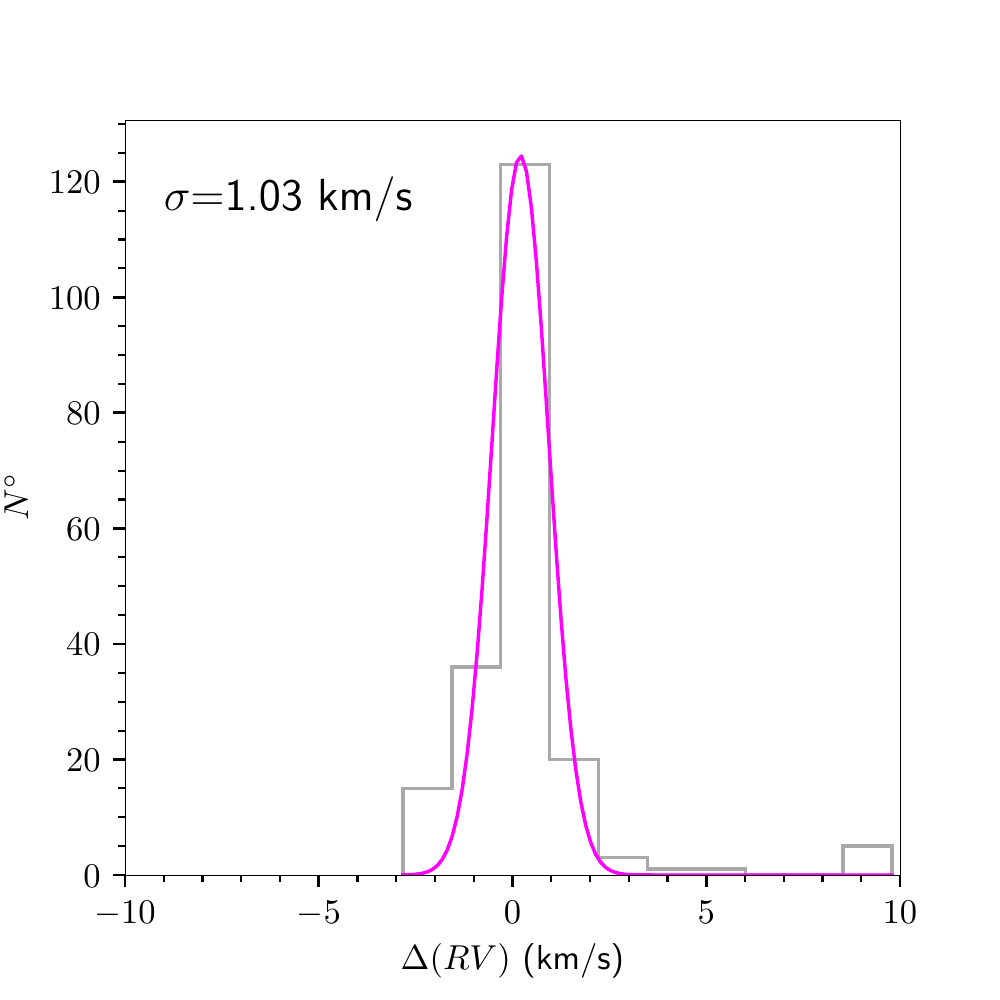}
\caption{Histogram of the differences---divided by the root square of 2---between pairs of radial velocities measurements for the same star.  The best-fitting Gaussian to the distribution is plotted, and its standard deviation $\sigma$  is  indicated.  }
    \label{fig:err_rv}
\end{figure}

We considered the errors returned by \texttt{fxcor} as conservative estimates of the true uncertainties of the radial velocity. For each star we have four to eight radial velocity measurements and \texttt{fxcor}  error estimations. We followed the same procedure of \citetalias{Rain_2020}, i.e., we computed first the \texttt{fxcor} error for each star, and for each pair of measurements we calculated the radial velocity difference divided by the root square of 2; then we built the distribution histogram and fitted a Gaussian. We considered the standard deviation $\sigma$ of the Gaussian as the true radial velocity error. We plotted the histogram together with the Gaussian fit and the true error in  Figure \ref{fig:err_rv}.  

Additionally, we calculated the mean \texttt{fxcor} error for each rotational rate, estimated as we described above in \S~\ref{subsec:rv}.  Stars rotating with velocities ranging between approximately  30 and 50~km~s$^{-1}$ have errors about 2--4~km~s$^{-1}$, and stars rotating with velocities of 80--150~km~s$^{-1}$ have uncertainties of around 10~km~s$^{-1}$; finally, the typical uncertainties for the fast rotator stars ($v \sin i > 150$~km~s$^{-1}$) are about 15~km~s$^{-1}$.  Similar uncertainties values were found by \citet{Mucciarelli_2014} on their BSS sample.
Therefore, we decided to adopt the \texttt{fxcor} error as a conservative estimation for the radial velocity uncertainty.

\subsection{Membership and evolutionary status}\label{subsec:Member_evol_status}
 
By means of the comparison between the radial velocities we have measured for our BSS candidates, and the mean radial velocity of the clusters, we can try now to assess possible membership.  In what follows we will assume that BSS are the result of collisions, or that they are binary systems, with either relatively short periods (a few days or less),  or long ones (about 1000 days). Given the example of NGC~188, we expect most BSS to be rather long-period binaries \citep{Geller_2011, Davies_2015, Ivanova_2015}. We adopted as the mean velocity of each cluster, that obtained in previous studies of clump stars.

For each cluster, we have between four and eight radial velocities, obtained at epochs separated by days, months, and years  (Table \ref{tab:obs_details}). To assess membership,  the radial velocities of the stars can be compared with the mean radial velocity of the cluster, taking into account the error bars---as derived in \S~\ref{subsec:rv}---and the possibility of binarity.

In this work we adopted the same criteria as in \citetalias{Rain_2020}.  
Statistical analysis of the mass distribution of the secondary components of BSS in binaries with orbital periods near 1000 days shows that the masses of the companions are peaked between 0.5 and 0.6~$M_{\odot}$ \citep{Geller_2011}. Such masses suggest white dwarf (WD) secondaries, whose presence would indicate mass transfer as the dominant formation mechanism in BSS. Using the TO mass values of our clusters  (1.12 to 1.5~$M_{\odot}$), and assuming the mass ratio $q$ of the systems to be 0.5, for the system not to fill its Roche lobe the separation between the stars should be larger than $\sim 3.5~R_{\odot}$. The minimum orbital period should be of about 0.5~days, with a corresponding maximum orbital velocity of about 100~km~s$^{-1}$. Therefore, stars with radial velocities changing between epochs up to 100~km~s$^{-1}$ from the cluster mean could still be considered as members, provided they are close binaries. On the other hand, if we consider post-mass-transfer, long-period binaries ($P\sim1000$ days), we would expect a maximum radial velocity of 10--13~km~s$^{-1}$. Of course, it is possible to have a binary system in between the two cases.

These considerations led us to define the following, rather conservative approach to confirm membership of BSS in our clusters.  If the individual radial velocities are, given their error bars, compatible with the cluster mean $V_{R}$, and do not change significantly over the four epochs, the star  is taken as a possible single-star member, i.e., the outcome of a collision or a merger. Of course, it could  also be a binary with a long period---larger than $\sim 1000$ days. These stars are classified \textbf{``M''}.  If a star's  velocity is compatible with the cluster mean $V_{R}$, but the velocity errors are too large (say, larger than 1/3 the radial velocity value)  to discriminate between long-period or close binaries, we indicate it as \textbf{``M?''}. ii) On the other hand, if the individual radial velocities are, given their error bars, within 100~km~s$^{-1}$  respect to the cluster mean $V_{R}$, then: a) If the velocities differ more than 20~km~s$^{-1}$ from $V_{R}$ and  change significantly between two epochs, we can consider the star as a candidate for being a close-binary member of the cluster, \textbf{``M, CB''}. When this happens, but error bars are too large, we tag the star as \textbf{``M, CB?''}. However, note that when the period is close to the difference in time between the epochs, we should not expect much change in radial velocity.
b) If the velocities are within 20 km~s$^{-1}$ from $V_{R}$ and do not change by more than a few km~s$^{-1}$ between epochs (depending on the possible period, which is constrained by the difference with $V_{R}$), we  possibly have a long-period (above 100 days) binary, and it is classified as  \textbf{``M, LP''}. When this is true, but error bars are too large, the classification is \textbf{``M, LP?''} instead.

The membership status of the binaries (CB and LP) can only be secured once we have determined the full orbital solution, and thus derived the systemic velocity.  If none of the above apply, we label the star as a non-member, \textbf{``NM''}.

\subsection{Spectroscopic detections}\label{sec:spec_detection}

Several explanations have been proposed for the blue straggler phenomenon, although none is completely satisfactory (cf.~\S~\ref{sec:introduccion}). The binarity hypothesis is the most accepted one, mainly because it can in principle account for most of the observations. \cite{Mathieu_2009} have shown that the percentage of binaries among BSS is significantly larger than in the cluster main sequence (MS). Previous studies in OCs had revealed that the BSS population in open clusters mostly contains long-period binaries  \citep{Geller_2009}; these have periods ranging from a few years to decades or even centuries, and it is very difficult to detect them spectroscopically and photometrically.  On the other hand,  \citetalias{Rain_2020} found a significant amount of BSS in  Collinder~261  being possible close binaries. 

Following the description of \S~\ref{subsec:Member_evol_status}, and based on their radial velocity variations, we attempted to roughly assess the binary nature of the sample studied here, namely, to decide if they may be close, long-period binaries, or single stars without radial velocity variations. All the probable binaries would need additional spectroscopic follow-up to be properly characterized, given the small number of observations.

\subsubsection{Trumpler 5} This object contains a larger number of stars than many old open clusters in the Galaxy. Despite its distance relatively close to the sun ($\sim 3$~kpc), given its location in a highly and differentially reddened region, not many photometric and spectroscopic studies have been carried out on their members, and none on its blue straggler population.
 
Out of our 40 BSS candidates, only seven were observed with FLAMES, for which five epochs of spectra were available.  Among these stars, we found four close binaries. We classified stars Tr5-BS2 (80\%), Tr5-BS3 (90\%), Tr5-BS6 (100\%), and Tr5-BS7 (100\%)  as members and possible close binaries (M, CB). On the other hand, we found three possible interlopers:  Tr5-BS1 (80\%),  Tr5-BS4 (100\%), and Tr5-BS5 (100\%) which, according to our criteria and measured radial velocities, are not members. The high probability of membership ($P_{\mathrm{memb}}$) that this three receive from \citetalias{Cantat-Gaudin_2018}  shows that a good astrometric solution---like that of \emph{Gaia} DR2---is not enough for a correct identification of a bona~fide blue straggler, but that spectroscopic data are also needed.
We found that almost all  stars in Trumpler~5 observed with FLAMES are fast rotators (including the interlopers),  except  Tr5-BS2. The theoretical expectations for  rotation velocities of BSS are not well defined, but all current scenarios can plausibly spin up them. These stars are similar to those  found by  \cite{Mucciarelli_2014}, with the fastest rotating up to $\sim200$ km~s$^{-1}$.

\subsubsection{Trumpler 20} Given its position in the inner disk (where not many old OCs reside), age, proximity, and mass,  Trumpler~20 is particularly interesting. The blue straggler population of this cluster has been identified in a handful of photometric studies (AL95, \citetalias{Ahumada_2007}, \citealt{Carraro_2010}), but not with spectroscopy. Unfortunately, only one star out of the eight BSS with $P_{\mathrm{memb}} \geq 50$\% was observed with FLAMES. Star Tr20-BS1 (100\%) is a possible close binary according to our criteria.   
Two stars with probabilities below 50\% were also observed, namely Tr20-BS2 (10\%) and Tr20-BS3 (30\%), which were identified as one possible close binary (member) and a non-member (possible long-period binary), respectively. For this cluster, we measured radial velocities in four epochs for the three stars.

\subsubsection{NGC 2477} This cluster is moderately old and slightly metal-poor. It is also known to be one of the clusters with the largest amount of member stars in the southern sky \citep{Gao_2018}. For the five BSS we identified in the CMD (bottom panels of Figure \ref{fig:cmd}), we obtained eight epochs of radial velocities for three stars: NGC2477-BS1 (100\%), NGC2477-BS2 (90\%), and NGC2477-BS3 (80\%). Despite the high binary frequency ($\sim 36$\%) found by \cite{Eigenbrod_2004},  our radial velocity measurements indicate that these three stars are single non-variable members---although they could also be long-period binaries or binaries seen face-on. If we look into their projected rotational rates, BSS do not seem to rotate unusually fast,  similarly to what \cite{Smith_1983} found. In the work previously mentioned, NGC2477-BS2 (90\%) or HART 7302 is classified as a G3IV-V dwarf and is considered to be a probable interloper according to its position in the CMD, allegedly close to the cluster TO. However, in our CMD  this object appears clearly separated from the TO and, given its proper motion, parallax, and radial velocities values, this star may be considered a member.  \cite{Smith_1983} give a rotational velocity of 50 km~s$^{-1}$  for this star, very close to the 45 km~s$^{-1}$ we found.

\section{Radial density profiles}
\label{sec:radial_density_profiles}

The radial distribution is a powerful tool to estimate the dynamical age of star clusters and has been extensively studied (e.g., \citetalias{Ferraro_2012} and \citealt{Beccari_2013}). In this context, and given their relatively high masses and luminosities, BSS are the perfect objects to analyse  dynamical friction and the mass segregation, in view of their direct relation with the cluster dynamics.

An analysis of the stellar radial distribution can only be carried out in Trumpler~5, given its
 significant number of blue stragglers.
To analyse the behaviour of BSS relative to that of normal stars, and to better understand the dynamical state of the cluster, we studied the BSS radial distribution and compared it to that of a reference population, assumed to trace the normal cluster stars. A bright, natural reference population in the optical is the red giant branch (RGB); given the accurate astrometric solutions of \emph{Gaia},  these stars can be very reliably identified. 
Sources with $G < G_{\mathrm{TO}}-0.5$ (in the de-reddened CMD) and $P_{\mathrm{memb}}\geq 50$\% were selected as RGB stars.
We thus identified 142 RGB stars in Trumpler~5. \\ 

\begin{figure*}
\includegraphics[width=\linewidth]{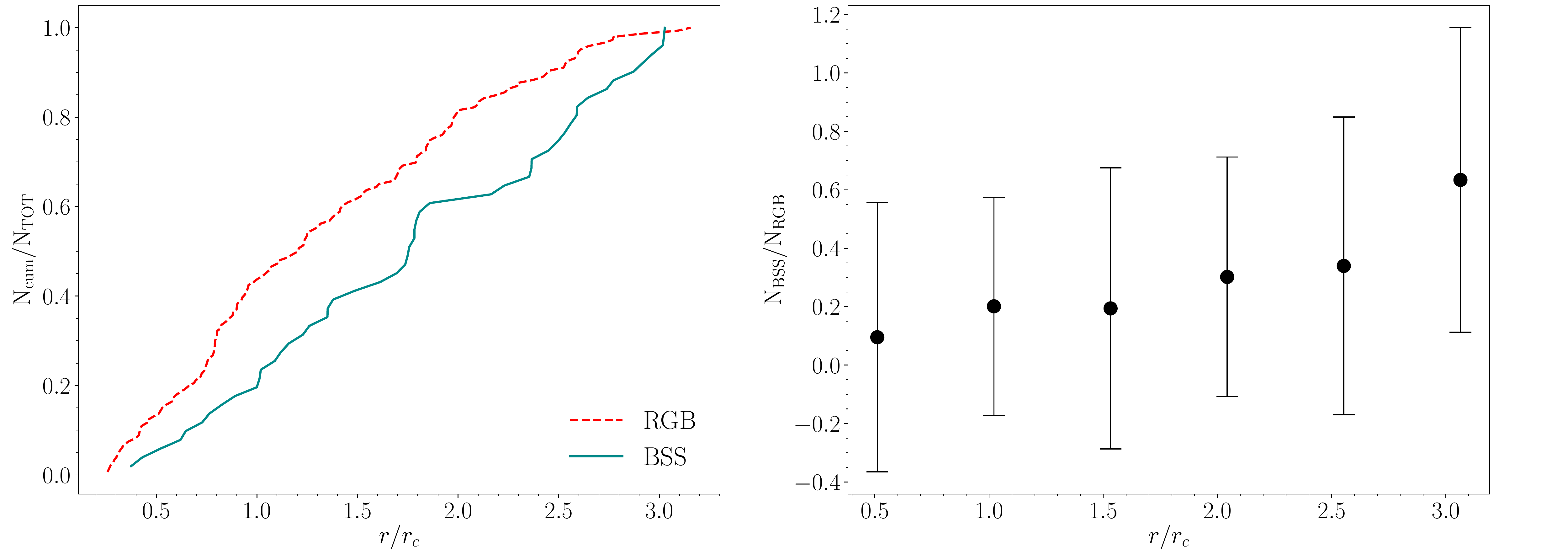}
\caption{\textbf{Left:} Cumulative spatial distribution of BSS (blue line) and RGB stars (red dashed line) in Trumpler~5. \textbf{Right:} Relative number of BSS to RGB stars, plotted as a function of the distance from the cluster center.  Errors are Poisson distributed. }
    \label{fig:radialdistribution-trump5}
\end{figure*}

\subsection{Cumulative Radial Distribution and Population Ratios}
\label{subsec:cumulative_ratios}

The cumulative spatial distributions on the $y$-axis as a function of $r/r_{c}$ is shown in the left panels of Figure~\ref{fig:radialdistribution-trump5}. The black solid line indicates the normalised cumulative distribution of the BSS candidates in comparison with the sample of RGB stars (red dashed line). For Trumpler~5 the BSS do not appear more centrally concentrated than the reference population. Our finding disagrees with what is observed in other clusters, whose BSS show high concentration in the cluster internal region relative to the evolved stars (\citealt{Geller_2008, Bhattacharya_2019, Vaidya_2020}, \citetalias{Rain_2020}). 

To quantify whether the radial distributions of BSS and RGB stars are extracted from the same parent distribution, thus indicating the absence of segregation, we used the $k$-sample Anderson-Darling test \citep[hereafter A-D test]{Scholz_1987}. The A-D test indicates a difference of 99.9\% between the distributions of BSS and RGB stars, i.e.,  both populations do not originate from the same distribution.

Additionally, a further indicator of segregation is the number of BSS normalised to the number of a reference population.  We divided the field of view in concentric annuli, such that each annulus has at least one BSS. The right panel of  Figure~\ref{fig:radialdistribution-trump5} show the number of BSS candidates with respect to that of  RGB stars in each annulus, as a function of $r/r_{c}$ . The ratio was corrected assuming that the field contamination we found in \S~\ref{subsec:field_contamination} is homogeneous. In the case of Trumpler~5---and considering the errors---the distribution does not become flat, indicating that BS and RGB stars are more or less equally distributed both  in the central part of the cluster and in its  outskirts.

Comparing this distribution with those described by \citetalias{Ferraro_2012} for globular clusters, it appears to be somewhat similar to that of Family~I clusters. In this family, the radial distribution of the stragglers is fully consistent with that of the reference population, and dynamical friction has not yet played a major role,  even in the core.

Following \citetalias{Vaidya_2020}, we decided to test the possible location of $r_{\mathrm{min}}$/$r_{c}$ in the radial distribution in this cluster. To do so, we first estimated the central relaxation time $t_{\mathrm{relax,c}} \sim t_{\mathrm{cross}}\, N_{*}/6\log N_\ast$ for the three clusters---for the sake of completeness, $t_{\mathrm{cross}} \sim D/\sigma_{v}$ is the crossing time, $N_{\ast}$ is the total number of stars (within the radius of each cluster and with  $P_{\mathrm{memb}}\geq 50$\%), and $\sigma_{v}$ is the velocity dispersion \citep{Binney_1987}. In the calculations we employed the standard deviations of the projected proper motions of each cluster, as well as  the core radius we derived from of King profiles (\S~\ref{subsec:king_profile}). Second, using the values of $t_{\mathrm{relax,c}}$ and the evolutionary age of each cluster, we estimated the parameter $N_{\mathrm{relax}} = \mathrm{age}/t_{\mathrm{relax,c}}$. These three parameters and the number of stars are reported in Table~\ref{tab:structural_parameters}. Finally, using the value of $N_{\mathrm{relax}}$ we estimated $r_{\mathrm{min}}/r_{c}$ using equation~(1) from \citetalias{Vaidya_2020}, but only for Trumpler~5.  The dynamical state of this cluster will be discussed in more detail in the Conclusions (\S~\ref{sec:conclusions}).

\section{Summary and Conclusions}\label{sec:conclusions}

We have studied the blue and yellow straggler population of the open clusters NGC~2477, Trumpler~5, and Trumpler~20  using the selection of cluster members published by \cite{Cantat-Gaudin_2018} and updated by \cite{Cantat-Gaudin_2020}. They performed a membership selection based on \emph{Gaia} latest release (\emph{Gaia}~DR2) astrometric solution. Additionally, we complemented our results with spectroscopic data from FLAMES/GIRAFFE and  defined a rough classification of their binary nature based on their radial velocity variability. The accuracy and precision of \emph{Gaia}~DR2 enable the discovery of astrometric binaries which are, however, significantly affected by the measurements and data processing. The \emph{Gaia} single body 5-parameter astrometric solution is fitted to the binary, under the assumption that the binary moves like a single point mass which leads to considerable biases. In this sense, we are aware that \citetalias{Cantat-Gaudin_2018} astrometric selection of members likely leaves out some binaries.  Long-period binaries will be affected by an excess of proper motions, and systems with periods around 1 year will have an effect on their parallax values (under- or over-estimation, \citealt{Penoyre_2020}). On the other hand, in close-binary systems the proper motions will be affected by the orbital velocity.

Using red clump stars and members with radial velocity measurements ($G<13$) in \emph{Gaia}, we calculated the cluster mean proper motions and  parallax; we found  large differences with the values reported by DAML02, being Trumpler~20 the worst  case with differences of $\sim4$--5~mas~yr$^{-1}$. We believe that these differences are not caused by systematic errors in \emph{Gaia}, but by the lack of reliable cluster membership and the presence of field contamination.

Before the selection of the straggler candidates, we corrected every CMD by the effect of differential reddening, and estimated the field contamination caused by young stars. The most affected cluster in both aspects is Trumpler~5, followed by Trumpler~20, not unsurprisingly  given their positions low onto the Galactic plane. In NGC~2477 the effect of both extinction and contamination is negligible. Of the three clusters, Trumpler~5 is the one that hosts the largest sample of blue stragglers candidates. We found 40 blue stragglers and three yellow stragglers among the cluster members (Table~\ref{tab:BS_phot_Tr5}). In the case of Trumpler~20, we identified only eight BS candidates (Table~\ref{tab:BS_phot_Tr20}), and in NGC~2477, five BSS and four YSS candidates were visible in the CMD   (Table~\ref{tab:BS_phot_NGC2477}). All YS candidates are listed in Table~\ref{tab:BS_phot_ys}, and the final results for the three clusters are in Figure~\ref{fig:cmd}. We found large inconsistencies in \citetalias{Ahumada_2007} blue straggler candidates, especially in Trumpler~5, where only $\sim$6 \% of the BSS listed in \citetalias{Ahumada_2007} are members. In the case of NGC~2477,  all \citetalias{Ahumada_2007} BSS are members but they are all concentrated around the TO. Evidently, the  quality of \emph{Gaia} data contributes to improve the construction of better, bona~fide lists of blue straggler candidates.
  
Following \cite{Bhattacharya_2019}, \citetalias{Rain_2020}, and \citetalias{Vaidya_2020}, we used our candidates as test-particles to probe the dynamical state of Trumpler~5. Our goal was to explore the bimodal distribution of BSS in open clusters, which is poorly understood, unlike its GC counterpart. First, we compared BSS candidates with a reference population (RGB stars)---expected to follow the cluster light distribution---selected within the de-reddened $G < G_{\mathrm{TO}}-0.5$ magnitude range; exploring the normalized cumulative radial distributions, we found that BSS are not more centrally concentrated than RGB stars (left panels of Figure \ref{fig:radialdistribution-trump5}) in Trumpler~5. Second, we plotted the ratio of BSS to RGB stars $N_{\mathrm{BSS}}/N_{\mathrm{RGB}}$ (see right panel of Figure \ref{fig:radialdistribution-trump5}); to do so, we split the field of view in concentric annuli, each one containing at least one BS. Before the comparison,  we performed an A-D test to check that both populations were not extracted from the same parent distribution.  The test gave a 99.9\% probability that RGB stars are not drawn from the same distribution as BSS. 
Based on his flat radial distribution, Trumpler~5 can be classified as a Family~I-type cluster.  Additionally, we calculated the predicted $r_{\mathrm{min}}$ using the correlation (Eq.~1) of \citetalias{Vaidya_2020} and the corresponding values of $N_{\mathrm{relax}}$ 
(Table~\ref{tab:structural_parameters}),  and obtained a value of $2.29\, r_{c}$.  The distribution of Trumpler~5 is flat up to  15~arcmin,  with no clear signs of a minimum in the BSS radial distribution at the predicted  $r_{\mathrm{min}}$. The case of Trumpler~5 is similar to both Berkeley~39 and NGC~6819 (see \citetalias{Vaidya_2020}), whose estimated values of $N_{\mathrm{relax}}$ suggest that they are dynamically evolved, but that their radial distributions are flat. 
In \S~\ref{sec:spec-analysis} we presented the first high-resolution spectroscopic study of the  BSS population of Trumpler~5, adopting more solid membership criteria  than the simple photometric ones.  For this cluster we obtained four epochs of radial velocities; based on their variations, we separated these stars into candidate members, probable close binaries, and long-period binaries. Unfortunately,  these data only cover seven out of the 40 possible BSS found in our analysis with \emph{Gaia}.  Our spectroscopic results are reported in Table~\ref{tab:rv_blue_Tr5}.  Radial velocities for four epochs are available for seven stars,  among which we identified four as probable contact binaries, and three as non-members. All the stars are fast rotators, with the exception of star Tr5-BS2. We conclude that Trumpler~5 hosts 37 blue straggler candidates within $r/r_{c} \sim 3.28'$.

Our spectroscopic results for Trumpler~20 are reported in Table \ref{tab:rv_blue_Tr20}. Radial velocities for five epochs are available for three stars. Only one of these stars was safely classified as a BS candidate (within the radius $R$ and with $P_{\mathrm{memb}} \geq 50$\%); this star is a possible close binary system. The remaining two stars with probabilities below 50\% are Tr20-BS2 (10\%) and Tr20-BS3 (30\%), identified as one possible close binary (member) and a non-member, respectively. Given our conservative criteria, it is expected that we would miss some genuine stragglers, and star Tr20-BS2 is an example. Although we did not attempt to estimate the number of uncounted BSS, we are confident that it  is small since our choice of members with $P_{\mathrm{memb}} \geq 50$\% captures the majority of the cluster members, as  has also been found by other authors. Having said that, we found Trumpler~20 hosts nine BSS.

In the case of NGC~2477, our results are shown in Table \ref{tab:rv_blue_NGC2477}. Radial velocities for eight epochs are available  for three stars, and only one of them has probabilities high enough to be a BS candidate.

We estimated the field contamination in all three cluster areas within their radius $R$ (Table~\ref{tab:clusters_parameters} and \S~\ref{subsec:field_contamination}) and  found 13 and 2 interlopers for Trumpler~5 and Trumpler~20, respectively, and none for NGC~2477. This means  that we have $\sim 25$\% of field contamination in the case of Trumpler~5 and $\sim 13$\% for Trumpler~20. Although this work shows an improvement over the photometric selection of \citetalias{Ahumada_2007}, the contamination in both clusters is quite high and consistent with the results from our spectroscopic follow-up, where 3/7 stars observed were classified as non-members in Trumpler~5. This shows, once again, that  such a spectroscopic follow-up is rather essential to a proper understanding of these exotic populations in OCs.

This study and  other recent ones like \citetalias{Rain_2020},  \cite{Bhattacharya_2019}, or \citetalias{Vaidya_2020}, indicate that there is an increasing interest for BSS in Galactic clusters. Besides, the evidence is emerging that a proper membership assessment is mandatory before performing any statistical analysis of the BSS population in open clusters.  We hope to be able to extend this membership assessment to many more clusters in the near future.

\section{Acknowledgements}
We are grateful to the anonymous referee for helpful comments which significantly helped improve the paper.\\
MJ.~R is supported by CONICY PFCHA through Programa de Becas de Doctorado en el extranjero- Becas Chile / \texttt{2018-72190617}.
\\S.~V.\ gratefully acknowledges the support provide by Fondecyt reg. \texttt{n. 1170518.} \\
J.~A.\ wishes to thank ESO for stays in the Santiago Headquarters in January 2010, November 2011, and October 2015, where part of this work was originally developed.
\\ This work is based on data obtained through ESO programs 088.D-0045(A) and 0100.D-0052(A).

\software{IRAF (Tody 1986, Tody 1993), UPMASK (Krone-Martins \& Moitinho 2014), Topcat (Taylor 2005), SPECTRUM (Gray \& Corbally 1994), ATLAS9 (Castelli \& Kurucz 2003)}

\bibliography{sample63}{}
\bibliographystyle{aasjournal}

\begin{table*}
\centering
\caption{Yellow straggler candidates from Gaia DR2 data for three open clusters.}
\begin{tabular}{cccccccccc}
\hline \hline
 \emph{Gaia} DR2 Source Id.  & $G_{corr}$ & $(G_\mathrm{BP}-G_\mathrm{RP})_{corr}$  & $\mu_{\alpha}\cos \delta$ &  $ \Delta \mu_{\alpha}\cos \delta$  &
 $\mu_{\delta}$  & $\Delta \mu_{\delta}$  & $\varpi$ & $\Delta \varpi$ & $P_{\mathrm{memb}}$  \\
 & [mag] & [mag] & [mas~yr$^{-1}$] & [mas~yr$^{-1}$] & [mas~yr$^{-1}$] & [mas~yr$^{-1}$] & [mas] & [mas] & \\
\hline
NGC~2477 & & & & & & & & & \\
  5538866344059037952 & 10.851 & 0.98 & -2.318 & 0.04 & 0.982 & 0.05 & 0.6276 & 0.0261 & 0.8\\
  5538869956136673792 & 10.564 & 1.245 & -2.54 & 0.051 & 0.938 & 0.062 & 0.6881 & 0.0315 & 1.0\\
  5538494679075370752 & 11.153 & 1.375 & -2.404 & 0.043 & 0.835 & 0.057 & 0.6368 & 0.027 & 0.7\\
  5538493579557251072 & 10.714 & 1.318 & -2.53 & 0.045 & 0.665 & 0.062 & 0.6548 & 0.0313 & 0.9\\
  \hline
Trumpler~5  & & & & & & & & & \\ 
   3326780722869330304 & 14.85 & 1.300 & -0.691 & 0.072 & 0.022 & 0.066 & 0.2245 & 0.0404 & 0.8\\
  3326841436526458496 & 14.621 & 1.362 & -0.179 & 0.088 & 0.528 & 0.075 & 0.0678 & 0.0465 & 0.5\\
  3326823603822996608 & 14.387 & 1.398 & -0.873 & 0.069 & 0.291 & 0.067 & 0.117 & 0.0371 & 1.0\\
  3326783746526306176 & 14.519 & 1.375 & -0.58 & 0.063 & 0.294 & 0.058 & 0.2805 & 0.0342 & 1.0\\
  3326809997366180608 & 14.795 & 1.353 & -0.501 & 0.073 & 0.333 & 0.062 & 0.3609 & 0.0462 & 1.0\\
  3326852328563919744 & 14.532 & 1.279 & -0.591 & 0.065 & 0.44 & 0.058 & 0.2445 & 0.0354 & 1.0\\
 \hline
 \label{tab:BS_phot_ys}
\end{tabular}\\
\end{table*}

\begin{deluxetable*}{ccccccccccc}
\tablecaption{Blue straggler candidates in NGC~2477 from \emph{Gaia}~DR2 data.\label{tab:BS_phot_NGC2477}}
\tablehead{
\colhead{\emph{Gaia} DR2 Source Id.}  & \colhead{$G_{corr}$} & \colhead{$(G_\mathrm{BP}-G_\mathrm{RP})_{corr}$}  & \colhead{$\mu_{\alpha}\cos \delta$} &  \colhead{$ \Delta \mu_{\alpha}\cos \delta$}  & 
 \colhead{$\mu_{\delta}$}  & \colhead{$\Delta \mu_{\delta}$}  & \colhead{$\varpi$} & \colhead{$\Delta \varpi$} & \colhead{$P_{\mathrm{memb}}$} & M \\
  & \colhead{(mag)} & \colhead{(mag)} & \colhead{(mas~yr$^{-1}$)} & \colhead{(mas~yr$^{-1}$)} &\colhead{(mas~yr$^{-1}$)} & \colhead{(mas~yr$^{-1}$)} & \colhead{(mas)} & \colhead{(mas)} & &
  }
\startdata
  5538883317769763072 & 12.364 & 0.437 & -2.493 & 0.041 & 0.813 & 0.046 & 0.7239 & 0.0252 & 0.9 &N\\
  5538871193083003264 & 11.561 & 0.454 & -2.329 & 0.055 & 0.85 & 0.057 & 0.696 & 0.0324 & 0.8 & N\\
  5538866722016153088 & 12.371 & 0.529 & -2.277 & 0.047 & 0.339 & 0.06 & 0.6589 & 0.0293 & 1.0 & N\\
  5538493682642912256 & 11.357 & 0.388 & -2.316 & 0.045 & 0.447 & 0.06 & 0.6468 & 0.0313 & 1.0 & N\\
  5538494885233812864 & 11.732 & 0.384 & -2.406 & 0.045 & 1.063 & 0.048 & 0.6434 & 0.0254 & 0.9 & N\\
  \enddata
 \end{deluxetable*} 



\startlongtable
\begin{deluxetable*}{ccccccccccc}
\tablecaption{Blue straggler candidates in Trumpler~5 from Gaia DR2 data.\label{tab:BS_phot_Tr5}. The last column indicates if the star is a massive candidate.}
\tablehead{
\colhead{\emph{Gaia} DR2 Source Id.}  & \colhead{$G_{corr}$} & \colhead{$(G_\mathrm{BP}-G_\mathrm{RP})_{corr}$}  & \colhead{$\mu_{\alpha}\cos \delta$} &  \colhead{$ \Delta \mu_{\alpha}\cos \delta$}  & 
 \colhead{$\mu_{\delta}$}  & \colhead{$\Delta \mu_{\delta}$}  & \colhead{$\varpi$} & \colhead{$\Delta \varpi$} & \colhead{$P_{\mathrm{memb}}$} & M \\
  & \colhead{(mag)} & \colhead{(mag)} & \colhead{(mas~yr$^{-1}$)} & \colhead{(mas~yr$^{-1}$)} &\colhead{(mas~yr$^{-1}$)} & \colhead{(mas~yr$^{-1}$)} & \colhead{(mas)} & \colhead{(mas)} & &
  }
\startdata
  3326766596718502400 & 15.444 & 0.865 & -0.97 & 0.12 & 0.561 & 0.094 & 0.3045 & 0.0566 & 0.8 & N\\
  3326770307572118144 & 15.1 & 0.83 & -0.735 & 0.069 & 0.819 & 0.058 & 0.3105 & 0.0418 & 1.0 & N\\
  3326766665438019968 & 15.117 & 0.791 & -0.756 & 0.081 & 0.41 & 0.067 & 0.4321 & 0.0469 & 0.9 & N\\
  3326774430739014144 & 15.967 & 0.769 & -0.804 & 0.119 & 0.639 & 0.12 & 0.3027 & 0.063 & 1.0 & N\\
  3326766974675728896 & 15.025 & 0.922 & 0.119 & 0.079 & 0.112 & 0.065 & 0.5137 & 0.0463 & 0.8 & N\\
  3326781616222424576 & 16.145 & 0.865 & -0.76 & 0.113 & 0.407 & 0.101 & 0.2707 & 0.0611 & 1.0 & N\\
  3326779726436867456 & 14.074 & 0.875 & -0.554 & 0.058 & 0.119 & 0.052 & 0.2909 & 0.0306 & 1.0 & N\\
  3326781203905606272 & 15.859 & 0.969 & -0.743 & 0.103 & 0.345 & 0.094 & 0.3361 & 0.0573 & 1.0 & N\\
  3326769624673711744 & 15.414 & 1.011 & -0.471 & 0.089 & 0.106 & 0.074 & 0.297 & 0.0489 & 1.0 & N\\
  3326777286895609344 & 13.522 & 0.55 & -0.244 & 0.052 & 0.032 & 0.055 & 0.4064 & 0.0247 & 0.8 & Y \\
  3326761412696403200 & 15.812 & 0.742 & -0.419 & 0.116 & -0.097 & 0.102 & 0.2973 & 0.0649 & 1.0 & N\\
  3326810680264949120 & 15.639 & 0.777 & -0.148 & 0.088 & 0.53 & 0.074 & 0.4389 & 0.0563 & 0.6 & N\\
  3326782303417071104 & 16.078 & 0.796 & -0.669 & 0.103 & 0.349 & 0.093 & 0.3731 & 0.0606 & 1.0 & N\\
  3326826043364268800 & 15.109 & 1.041 & -1.313 & 0.08 & 0.866 & 0.075 & 0.4134 & 0.0479 & 0.5 & N\\
  3326786254787049856 & 14.478 & 1.12 & -0.543 & 0.061 & 0.297 & 0.057 & 0.2738 & 0.0333 & 1.0 & N\\
  3326834736377867264 & 15.796 & 0.691 & -0.586 & 0.093 & 0.584 & 0.082 & 0.2038 & 0.0556 & 0.9 & N\\
  3326825459248778112 & 14.858 & 0.918 & -0.608 & 0.094 & 0.435 & 0.103 & 0.3007 & 0.0607 & 1.0 & N\\
  3326784154545235072 & 15.93 & 0.959 & -0.591 & 0.154 & 0.627 & 0.14 & 0.2261 & 0.0768 & 0.9 & N\\
  3326787079420759936 & 15.196 & 0.958 & -0.294 & 0.077 & 0.111 & 0.068 & 0.2874 & 0.0435 & 1.0 & N\\
  3326784399361287808 & 15.455 & 0.699 & -0.431 & 0.092 & 0.289 & 0.08 & 0.2367 & 0.0507 & 1.0 & N\\
  3326787487440281088 & 15.776 & 0.944 & -0.223 & 0.112 & 0.081 & 0.103 & 0.3073 & 0.0615 & 0.9 & N\\
  3326786082988414592 & 15.426 & 0.682 & -0.503 & 0.088 & 0.304 & 0.077 & 0.4483 & 0.0476 & 0.9 & N\\
  3326783746526299008 & 16.052 & 0.852 & -0.581 & 0.127 & 0.21 & 0.126 & 0.2884 & 0.0672 & 1.0 & N\\
  3326786426585687040 & 14.343 & 0.687 & -0.975 & 0.063 & 0.43 & 0.058 & 0.4 & 0.031 & 0.9 & N\\
  3326810478402457088 & 15.703 & 0.77 & -0.534 & 0.087 & 0.432 & 0.075 & 0.2583 & 0.0644 & 1.0 & N\\
  3326786594087018240 & 15.431 & 0.681 & -0.775 & 0.093 & 0.074 & 0.08 & 0.2791 & 0.0498 & 0.9  & N\\
  3326807557824686336 & 13.418 & 0.591 & 0.08 & 0.055 & 0.086 & 0.046 & 0.3445 & 0.027 & 0.5 & Y\\
  3326832984031309312 & 14.23 & 1.038 & -0.039 & 0.062 & -0.074 & 0.055 & 0.3185 & 0.0393 & 0.5 & N\\
  3326811096877800448 & 14.922 & 0.712 & -0.505 & 0.071 & 0.349 & 0.062 & 0.2993 & 0.0446 & 1.0 & N\\
  3326786903324593152 & 16.338 & 0.832 & -0.723 & 0.146 & 0.155 & 0.135 & 0.1964 & 0.0782 & 0.9 & N\\
  3326834598938944640 & 16.131 & 0.737 & -0.501 & 0.117 & -0.181 & 0.108 & 0.3239 & 0.0709 & 0.5 & N\\
  3326783368568993664 & 15.704 & 0.787 & -0.524 & 0.088 & 0.136 & 0.076 & 0.3578 & 0.0483 & 1.0 & N\\
  3326832606074193792 & 15.054 & 1.024 & -0.498 & 0.073 & 0.33 & 0.064 & 0.3076 & 0.0432 & 1.0 & N\\
  3326805908557292416 & 15.622 & 0.894 & -0.531 & 0.082 & 0.302 & 0.071 & 0.2298 & 0.0508 & 1.0 & N\\
  3326810306603807104 & 15.274 & 0.736 & -0.294 & 0.08 & -0.026 & 0.075 & 0.3726 & 0.0537 & 0.9 & N\\
  3326783570429587072 & 14.051 & 0.633 & -0.247 & 0.139 & -0.139 & 0.131 & 0.2848 & 0.0423 & 0.8 & N\\
  3326859956425708672 & 15.411 & 0.742 & -0.503 & 0.095 & 0.269 & 0.079 & 0.2726 & 0.0617 & 1.0 & N\\
\enddata
 \end{deluxetable*}

\begin{deluxetable*}{ccccccccccc}\label{tab:BS_phot_Tr20}
\tablecaption{Blue straggler candidates in Trumpler~20 from Gaia DR2 data. The last column indicates if the star is a massive candidate.}
\tablehead{
\colhead{\emph{Gaia} DR2 Source Id.}  & \colhead{$G_{corr}$} & \colhead{$(G_\mathrm{BP}-G_\mathrm{RP})_{corr}$}  & \colhead{$\mu_{\alpha}\cos \delta$} &  \colhead{$ \Delta \mu_{\alpha}\cos \delta$}  & 
 \colhead{$\mu_{\delta}$}  & \colhead{$\Delta \mu_{\delta}$}  & \colhead{$\varpi$} & \colhead{$\Delta \varpi$} & \colhead{$P_{\mathrm{memb}}$} & M \\
  & \colhead{(mag)} & \colhead{(mag)} & \colhead{(mas~yr$^{-1}$)} & \colhead{(mas~yr$^{-1}$)} &\colhead{(mas~yr$^{-1}$)} & \colhead{(mas~yr$^{-1}$)} & \colhead{(mas)} & \colhead{(mas)} & & 
  }
\startdata
 6056511543677927552 & 13.654 & 0.596 & -7.162 & 0.029 & 0.388 & 0.028 & 0.3259 & 0.0231 & 0.5& N\\
  6056526112167919232 & 15.757 & 0.811 & -7.095 & 0.048 & 0.153 & 0.048 & 0.2721 & 0.0337 & 1.0& N\\
  6056525425013827200 & 15.212 & 0.609 & -6.975 & 0.046 & 0.278 & 0.04 & 0.2123 & 0.0285 & 1.0& N\\
  6056527825849449344 & 14.672 & 0.641 & -7.062 & 0.036 & 0.021 & 0.034 & 0.2633 & 0.0262 & 1.0& N\\
  6056527417878920960 & 14.655 & 0.908 & -7.07 & 0.031 & 0.126 & 0.032 & 0.2258 & 0.0236 & 0.9& N\\
  6056530235377647872 & 15.611 & 0.811 & -7.114 & 0.049 & 0.104 & 0.048 & 0.2478 & 0.037 & 1.0& N\\
  6056570642433872512 & 15.514 & 0.794 & -7.211 & 0.045 & 0.029 & 0.039 & 0.198 & 0.0315 & 0.7& N\\
  6056577136424997376 & 14.572 & 0.719 & -7.08 & 0.03 & 0.281 & 0.029 & 0.2737 & 0.0219 & 1.0& N\\
  6056577548741835136 & 14.912 & 0.865 & -7.313 & 0.033  & 0.284 & 0.031 & 0.2590 &  0.0233 &  0.1 & N\\
   \enddata
 \end{deluxetable*} 



\end{document}